\documentclass[12pt]{article}
\pdfoutput=1
\usepackage[latin1]{inputenc}
\usepackage{amsmath}
\usepackage{amsfonts}
\usepackage{amssymb}
\usepackage{graphicx}
\usepackage{geometry}
\usepackage{amssymb,epsfig}
\usepackage{hyperref}
\usepackage{subfig}
\usepackage{amssymb, amsmath, amsopn, amsthm}
\usepackage{epsfig}
\usepackage{graphics}

\newcommand{\ti}{\tilde}

\makeatletter
\renewcommand\section{\@startsection {section}{1}{\z@}%
                                 {-3.5ex \@plus -1ex \@minus -.2ex}%nn
                                   {2.3ex \@plus.2ex}%
                                   {\normalfont\large\bfseries}}
\renewcommand\subsection{\@startsection{subsection}{2}{\z@}%
                                   {-3.25ex\@plus -1ex \@minus -.2ex}%
                                     {1.5ex \@plus .2ex}%
                                     {\normalfont\bfseries}}
\renewcommand\subsubsection{\@startsection{subsubsection}{3}{\z@}%
                                   {-3.25ex\@plus -1ex \@minus -.2ex}%
                                     {1.5ex \@plus .2ex}%
                                     {\normalfont\itshape}}
\makeatother

\def\pplogo{\vbox{\kern-\headheight\kern -29pt
\halign{##&##\hfil\cr&{\ppnumber}\cr\rule{0pt}{2.5ex}&\ppdate\cr}}}
\makeatletter
\def\ps@firstpage{\ps@empty \def\@oddhead{\hss\pplogo}%
  \let\@evenhead\@oddhead % in case an article starts on a left-hand page
}%      The only change in \maketitle is \thispagestyle{firstpage} instead of \thispagestyle{plain}
\def\maketitle{\par
 \begingroup
 \def\thefootnote{\fnsymbol{footnote}}
 \def\@makefnmark{\hbox{$^{\@thefnmark}$\hss}}
 \if@twocolumn
 \twocolumn[\@maketitle]
 \else \newpage
 \global\@topnum\z@ \@maketitle \fi\thispagestyle{firstpage}\@thanks
 \endgroup
 \setcounter{footnote}{0}
 \let\maketitle\relax
 \let\@maketitle\relax
 \gdef\@thanks{}\gdef\@author{}\gdef\@title{}\let\thanks\relax}
\makeatother
\makeatletter
\newcommand{\footnoteref}[1]{%
    \begingroup
        \unrestored@protected@xdef\@thefnmark{\ref{#1}}%
    \endgroup
    \@footnotemark}
\makeatother

\numberwithin{equation}{section}

\newcommand{\be}{\begin{equation}}
\newcommand{\bea}{\begin{eqnarray}}
\newcommand{\ee}{\end{equation}}
\newcommand{\eea}{\end{eqnarray}}
\newcommand\beq{\begin{equation}}
\newcommand\eeq{\end{equation}}
\newcommand{\f}{\frac}

\begin{document}

\setcounter{page}0
\def\ppnumber{\vbox{\baselineskip14pt
%\hbox{hep-th/0000000}
}}
\def\ppdate{\footnotesize{SU-ITP-12/39, TIFR/TH/12-45, YITP-12-92, WIS/19/12-DEC-DPPA}} \date{}

\author{Norihiro Iizuka$^1$, Shamit Kachru$^2$, 
Nilay Kundu$^3$, Prithvi Narayan$^4$,\\
\smallskip
 Nilanjan Sircar$^3$, Sandip P. Trivedi$^3$ and Huajia Wang$^2$\\
[7mm]
{\normalsize \it$^1$Yukawa Institute for Theoretical Physics}\\ 
{\normalsize \it Kyoto University, Kyoto 606-8502, JAPAN}\\
[5mm]
{\normalsize \it $^2$Stanford Institute for Theoretical Physics }\\
{\normalsize  \it Department of Physics, Stanford University}\\
{\normalsize \it Stanford, CA 94305, USA} \\
[1mm]
{\normalsize \it and}\\
[1mm]
{\normalsize \it Theory Group, SLAC National Accelerator Laboratory}\\
{\normalsize \it Menlo Park, CA 94309, USA}\\
[7mm]
{\normalsize \it $^3$Tata Institute of Fundamental Research}\\
{\normalsize \it Mumbai 400005, INDIA}\\
[7mm]
{\normalsize \it $^4$Weizmann Institute of Science}\\
{\normalsize \it Rehovot 76100, ISRAEL}\\
}

\bigskip
\title{\bf  Extremal Horizons with Reduced Symmetry: Hyperscaling
Violation, Stripes, and \\ 
 a Classification for the Homogeneous Case
 \vskip 0.1cm}%0.5cm}
\maketitle

\begin{abstract}

Classifying the zero-temperature ground states of quantum field theories with finite charge density is a very interesting problem. Via holography, this problem is mapped to the classification of extremal charged black brane geometries with anti-de Sitter asymptotics. In a recent paper \cite{Bianchi}, we proposed a Bianchi classification of the extremal near-horizon geometries in five dimensions, in the case where they are homogeneous but, in general, anisotropic. Here, we extend our study in two directions: we show that Bianchi attractors can lead to new phases, and generalize the classification of homogeneous phases in a way suggested by holography. In the first direction, we show that hyperscaling violation can naturally be incorporated into the Bianchi horizons. We also find analytical examples of ``striped" horizons. In the second direction, we propose a more complete classification of homogeneous horizon geometries where the natural mathematics involves real four-algebras with three dimensional  sub-algebras. This gives rise to a richer set of possible near-horizon geometries, where the holographic radial direction is non-trivially intertwined with field theory spatial coordinates. We find examples of several of the new types in systems consisting of reasonably simple matter sectors coupled to gravity, while arguing that others are forbidden by the Null Energy Condition. Extremal horizons in four dimensions governed by  three-algebras or four-algebras are also discussed. 

\end{abstract}
\bigskip
\newpage

\tableofcontents

\vskip 1cm

%%%%%%%%%%%%%%%%%%%%%%%%%%%%%%%%%%%%%%%%%%%
%%%%%%%%%%%%%%%%%%%%%%%%%%%%%%%%%%%%%%%%%%%
%%%%%%%%%%%%%%%%%%%%%%%%%%%%%%%%%%%%%%%%%%%
%%%%%%%%%%%%%%%%%%%%%%%%%%%%%%%%%%%%%%%%%%%
\section{Introduction}\label{sec:intro}

There has been considerable recent interest in possible applications of holography to condensed matter systems \cite{sean,chris,john,subir}.  
Some of the most interesting phenomena arise when one ``dopes" an insulator with finite charge density \cite{doping}.  Related fascinating phenomena
are thought to occur also in the phase diagram of QCD as one varies the chemical potential and number of light quark flavors \cite{RW}.
The problem of exemplifying interesting ground states of doped quantum field theory, and perhaps even classifying them, is then a difficult but well-motivated one.

For the subset of field theories with weakly curved gravitational duals, holography maps these complicated questions about quantum dynamics to
simple questions in classical general relativity.  Finite temperature states in the field theory are dual to black hole geometries 
with planar horizons (``black branes"), and states at
finite chemical potential (and charge density) map to charged black branes.  The low-temperature limit of such a system -- dual to the
ground state of the doped field theory -- is then governed by the extremal charged black brane geometry.  Therefore, in holography, the 
task is to present interesting examples of, or perhaps to classify, extremal black brane geometries in asymptotically anti de Sitter solutions of Einstein gravity coupled to various matter fields (chosen
to be typical of the content of low-energy string theory).

Studies of doped matter in AdS/CFT have yielded new near-horizon geometries with novel properties including dynamical scaling \cite{lifsol1,lifsol2,lifsol3}
and hyperscaling violation \cite{hv1,hv2,hv31,hv3,hv32,hv4,hv5,hv6}.  These are dual to emergent infrared phases which break Lorentz-invariance (as the doping does in the
UV), but which respect spatial rotation and translation invariance.
However, more typically, one expects to find emergent geometries which break more of the space-time symmetries.  In condensed matter physics,
for instance, phases with spin or charge density waves, stripe order, nematic order, and other more exotic orders are well known.  Similar modulated
phases also occur in the phase diagram of finite-density QCD.

There have been interesting holographic studies of some such spatially modulated phases -- notably, studies of emergent helical order \cite{harvey,MVR,ooguri,gauntlett}, stripe order \cite{holstripes1, Donosstripes, holstripes2,holstripes3, Rozali,DGnew}, and more elaborate orders \cite{vadim,jan,johanna}.  One largely open problem is to give simple analytical examples of phases realizing or even combining the various properties mentioned above (e.g. hyperscaling violation with
helical order) -- many cases discussed in the literature are complicated enough that numerical work is required.
 But a systematic classification of possible emergent geometries is also something one could hope to achieve.  That this is not always hopeless is clear from the success of the Kerr-Newman classification of charged, rotating black holes 
in asymptotically flat four-dimensional
space-time, and perhaps more relevant in this context, the success of the Bianchi classification of homogeneous cosmological solutions.
In fact, in a previous paper \cite{Bianchi}, we found that the basic strategy of the Bianchi classification in cosmology can also be used to classify
homogeneous (but, in general, anisotropic) extremal black brane geometries.  

The focus in \cite{Bianchi} was the symmetry structure of the 3d spatial
slices (the spatial dimensions ``where the field theory lives") in asymptotically $AdS_5$ space.  This gives rise to a classification
based on real three-dimensional Lie algebras, the original Bianchi classification \cite{Bianchioriginal} --
very readable expositions appear in \cite{Ryan, Landau}.  The generators of the Lie algebra correspond to Killing vectors in the geometry, which generate the isometry group of the corresponding spatial slices.\footnote{In \cite{Bianchi}, we also considered some examples 
where the three-algebra involves the time coordinate, which yield stationary space-times. }

In this paper, our goal is to apply and generalize the results of \cite{Bianchi} in several directions.  We show that by using standard ideas
of dimensional reduction and the solutions of \cite{Bianchi}, one can give simple analytical examples of  Bianchi horizons which exhibit hyperscaling violation.
Similarly, we also find analytical examples of ``striped" phases,  by using techniques of dimensional reduction.
And finally, we attempt to give a more general classification of
the possible homogeneous, anisotropic near-horizon metrics dual to doped 3+1 dimensional field theories.

The basic generalization of the classification of \cite{Bianchi} that remains possible can be easily explained.  Even with a focus on 
static solutions in 5d gravity, there is a more
general symmetry structure that could be relevant.  The additional ``radial" direction of holography could be non-trivially intertwined with
the spatial field theory coordinates.  This makes it clear that the relevant symmetry structure could more generally involve real four-dimensional
Lie algebras with a preferred three sub-algebra (which, roughly, generates the spatial symmetry group in the dual field theory).\footnote{The examples in \cite{Bianchi} are a degenerate case of this structure where there is a semi-simple Lie algebra with a separate Abelian factor corresponding to the isometry of translations in the radial direction.} 
Happily, the
classification of such algebras (with the relevant subalgebras) has also been accomplished, though much more recently -- see 
\cite{Patera} for a clear exposition and also \cite{MacCallum} for the proper history.  Here, we use
  these results to find the more general
classification of extremal black brane geometries relevant for holography in 5 bulk dimensions.   We also show that a few of the
new symmetry structures ${\it cannot}$ arise in solutions of theories that satisfy simple physical conditions, such as respecting the Null Energy Condition (NEC), and we give simple examples of
some interesting new types that do not run afoul of the NEC.

The more general possibilities described above also arise in the four dimensional spacetime,
 implying that one can find homogeneous horizons there
governed by real three-dimensional algebras with  two-dimensional  sub-algebras -- 
i.e., governed by the Bianchi types, but now with the symmetries acting on the ``radial" direction 
in addition to the ``field theory" spatial coordinates.  We can also potentially realize the real four-algebras 
in four dimensional  space-times. In such cases, the time-direction is nontrivially involved and as 
a result, we generically obtain either time-dependent metrics or stationary metrics, though static metrics occasionally arise.
We describe several such examples in the penultimate section. 

The organization of this paper is as follows.
%In \S2, we discuss basic background on the equations of motion in Einstein gravity coupled to a massive Abelian gauge fields. 
In \S \ref{sec:KKred}, we describe generic Kaluza-Klein dimensional  reduction of 
Einstein gravity coupled to massive vector fields. 
In \S \ref{sec:hypervioKK} - \S \ref{sec:hyperVII}, we discuss how to obtain
hyperscaling violating space-time metrics with reduced spatial symmetries.  
This is done by dimensional reduction of 5d Bianchi
metrics in \S \ref{sec:hypervioKK}, and by analyzing directly in 5d generalizations of the Bianchi types in \S \ref{sec:hyperVI} and \S \ref{sec:hyperVII}.
In \S \ref{sec:stripe}, we give examples of analytical solutions for ``striped"
phases, by Kaluza-Klein reducing the 5d type VII$_0$ geometries of \cite{Bianchi}.   
In \S  \ref{sec:Bianchi4algebras}, we change gears and discuss 
the four-dimensional real Lie algebras and 
their three-dimensional subalgebras.  
We also present invariant vector-fields and invariant one-forms which geometrically represent the algebra.
One might hope that one could realize all of these symmetry structures in holographic space-times, 
where the holographic radial direction is nontrivially intertwined with field theory spatial coordinates.  
However, in \S \ref{sec:NECfor5d}, we show that the Null Energy Condition (NEC) gives interesting constraints and
rules out several of the potentially relevant algebraic structures.
In \S \ref{sec:5d4algebra}, we show that prototypical examples of the remaining types enumerated 
in \S \ref{sec:NECfor5d} do arise as solutions of simple
gravity coupled to  matter theories.
In \S \ref{sec:Bianchi4dfor4d}, we describe 4d static space-times 
where the radial direction is nontrivially intertwined with ``field theory" space-time coordinates by 
either three-algebras or four-algebras and also  show how the NEC constrains or forbids some of these space-times.
We conclude in \S \ref{sec:discussion} with a brief discussion of possible future work.  
Some additional helpful  details are relegated to Appendices A-C.

%%%%%%%%%%%%%%%%%%%%%%%%%%%%%%%%%%%%%%%%%%%%%%%%%%%%%%%%%%%%%%%%%%%%%%%%%%%%%%%%%%%%%%%%%%%%%%%%%%%

\section{Kaluza-Klein reduction of gravity coupled to massive vectors}{\label{sec:KKred}}

In this section, we review the Kaluza-Klein reduction of gravity coupled to massive vector fields. 
We will use this result later at  \S \ref{sec:hypervioKK} and \S \ref{sec:stripe} in order to 
obtain hyperscaling violating and striped-phase metrics in lower dimensions.   

Consider the $d+1$ dimensional Einstein-Hilbert action coupled to massive  vector fields and a cosmological constant,
\begin{equation}~\label{EHaction_V_L}
 S=\frac{1}{\hat \kappa^2}\int d^{d+1} \hat x \sqrt{|\hat g|}\left[ \hat R
-\frac{1}{4} \hat F^{(i)2}-\frac{1}{4} m_i^2 \hat A^{(i)2} + \Lambda \right]
\end{equation}
where $\hat F^{(i)}= d \hat A^{(i)}$ is the field strength corresponding to the $i$-th massive vector field $A^{(i)}$, and where we note that our convention for $\Lambda$ requires positive $\Lambda$ to
support AdS space.
Quantities  with a $\hat ~$ correspond to fields in $(d+1)-$dimensions, while hatless objects are $d-$ dimensional.
 We now follow the calculation of \cite{KKLectPope}. 

Let us consider a split of the $d+1$-dimensions into $x^{\hat \mu}=(x^{\mu} ,z)$,
where $\hat \mu =1,\cdots, d+1$ and $\mu=1,\cdots,d$. We focus on field configurations where none of 
the $(d+1)$-dimensional fields depend on
the coordinate $z$, \emph{i.e.} $\partial_z$ is a Killing vector.
Consider the reduction ansatz to be:
\begin{eqnarray}~\label{ansatzg}
 d \hat s^2 &=& e^{2 \alpha_1 \phi(x)} ds^2+ e^{2 \alpha_2 \phi(x)} \left(dz+B_{\mu}(x) dx^{\mu}\right)^2 \,, \\
\hat A^{(i)}(x,z) &=& A^{(i)}_{\mu}(x) dx^{\mu}+ \chi^{(i)}(x) dz=\left(A^{(i)}(x)-\chi^{(i)}(x) B(x)\right)+ \chi^{(i)}(x) \left(dz+B\right) \,,
\end{eqnarray}
where $B=B_{\mu} dx^{\mu}$. The determinant of the metric satisfies,
\begin{equation}
 \sqrt{|\hat g|}=e^{(d \alpha_1+\alpha_2) \phi} \sqrt{| g|}~.
\end{equation}

The field strength is given by,
\begin{eqnarray}
 \hat F^{(i)} &=& d  \hat A^{(i)}=\left(dA^{(i)}-d\chi^{(i)}\wedge B\right)+ d\chi^{(i)}\wedge \left(dz+B\right) \\
&=& F^{(i)}+ d\chi^{(i)}\wedge \left(dz+B\right) \,,
\end{eqnarray}
where $F^{(i)}=dA^{(i)}-d\chi^{(i)}\wedge B$. 
The vierbein basis is related as,
\begin{equation}
 \hat E^a=e^{\alpha_1 \phi} E^a ~~;~~ \hat E^{\bar z}=e^{\alpha_2 \phi}\left(dz+B\right) \,,
\end{equation}
where $E^A$ is the vierbein in $(d+1)$-dimensions, and index $A=(a,\bar z)$.
Now (we will drop the index $i$ for convenience)
\begin{eqnarray}
 \hat F &=& \frac{1}{2} \hat F_{AB} \hat E^A \wedge \hat E^B \\
&=& \frac{1}{2} e^{2 \alpha_1 \phi} \hat F_{ab} E^a \wedge E^b+e^{( \alpha_1 +\alpha_2)\phi}\hat F_{a\bar z} E^a \wedge (dz+B)\\
&=& \frac{1}{2} F_{ab} E^a \wedge E^b+C_a  E^a \wedge (dz+B) \,,
\end{eqnarray}
where $C=d\chi$,
\begin{equation}
 \hat F_{ab}= e^{-2 \alpha_1 \phi} F_{ab} ~~;~~\hat F_{a\bar z}=e^{-(\alpha_1 +\alpha_2)\phi} C_a~.
\end{equation}
The components of the potential are related as
\begin{equation}
 \hat A_a =e^{- \alpha_1 \phi} (A-\chi B) ~~;~~ \hat A_{\bar z}= e^{- \alpha_2 \phi} \chi~.
\end{equation}

So the kinetic term of the gauge field gives
\begin{equation}
 -\frac{1}{4} \sqrt{|\hat g|} \hat F^2=-\frac{1}{4} \sqrt{|g|} e^{((d-4)\alpha_1+\alpha_2)\phi} F^2-
\frac{1}{2} \sqrt{|g|} e^{((d-2)\alpha_1-\alpha_2)\phi} C^2~.
\end{equation}
Similarly, the mass term satisfies
\begin{equation}
-\frac{1}{4} m^2 \hat A^2=-\frac{1}{4} m^2 e^{- 2 \alpha_1 \phi} (A-\chi B)^2-\frac{1}{4} m^2 e^{- 2\alpha_2 \phi} \chi^2~,
\end{equation}
so
\begin{equation}
-\sqrt{|\hat g|}\frac{1}{4} m^2 \hat A^2=-\frac{1}{4} m^2 e^{\left[(d- 2) \alpha_1 +\alpha_2\right]\phi} (A-\chi B)^2-\frac{1}{4} m^2 e^{(d\alpha_1- \alpha_2) \phi} \chi^2~.
\end{equation}

Now the gravitational part of the action reduces to~\cite{KKLectPope}
\begin{equation}
 \sqrt{|\hat g|} \hat R=\sqrt{|g|} \left( R-\frac{1}{2} (\partial \phi)^2-\frac{1}{4} e^{-2(d-1) \alpha_1 \phi} H^2 \right)~,
\end{equation}
where $H=dB$ and to write the gravitational action in the above form, we have used the identities
\begin{eqnarray}
\label{alpha_2alpha_1byd}
 \alpha_2 = -(d-2) \alpha_1 \quad \,,  \quad
\alpha_1^2 = \frac{1}{2 (d-1)(d-2)}~ \,.
\end{eqnarray}
These relations guarantee both that the Weyl rescaling puts the gravitational action in Einstein frame, and that the scalar field has a canonical
kinetic term.

In summary, the total dimensionally reduced action becomes:
\begin{eqnarray}~\label{dimredaction}
 S &=&\frac{1}{\kappa^2}\int d^{d} \hat x \sqrt{| g|}~
\lbrack  R -\frac{1}{2} (\partial \phi)^2-\frac{1}{4} e^{-2(d-1) \alpha_1 \phi} H^2
-\frac{1}{4} e^{-2 \alpha_1 \phi} F^{(i)2}-\frac{1}{2} e^{2 (d-2) \alpha_1 \phi} (\partial \chi^{(i)})^2 \nonumber \\
&& -\frac{1}{4} m_i^2 (A^{(i)}-\chi^{(i)} B)^2 -\frac{1}{4} m^2 e^{2 (d-1) \alpha_1
\phi} \chi^2 + \Lambda e^{2 \alpha_1 \phi}\rbrack~.
\end{eqnarray}

\section{Hyperscaling violation via Kaluza-Klein reduction}{\label{sec:hypervioKK}}

There has been considerable recent interest in the geometries that include a hyperscaling violation exponent $\theta \neq 0$
\cite{hv1,hv2,hv31,hv3,hv32,hv4,hv5,hv6}.  This is in part because, in some specific cases, they share many of the properties
of theories with a Fermi surface \cite{hv3}.  The cases studied in the literature have all enjoyed standard spatial rotation and
translation symmetries.  Here, we demonstrate that it is simple to find homogeneous
but anisotropic solutions which also enjoy
$\theta \neq 0$.  These can be duals to ordered phases with hyperscaling violation.

We actually do this in two steps.  As a first step, in this section, we show that dimensional reduction allows one to turn some of the 5d Bianchi
solutions into 4d Bianchi solutions with hyperscaling violation.  (By a 4d Bianchi solution, we mean a solution realizing one of the
Bianchi types with the radial direction included as one of the spatial dimensions governed by the algebra).  
In \S \ref{subsec:dimred}, we demonstrate this for type III solutions, and in \S \ref{subsec:typeIII} we show that the type III solutions of this sort can be obtained from
more general actions in 4d which do not arise from dimensional reduction.

As a next step, in \S \ref{sec:hyperVI}, we demonstrate that 5d Bianchi solutions (where the algebra acts only on the field theory ``spatial" coordinates)
can also enjoy hyperscaling violation.  We give several examples there, leaving the special case of type VII$_0$ (which has been especially
popular in the literature) to its own section, \S \ref{sec:hyperVII}.

Before proceeding  let us note 
that in the gravity theory hyperscaling violation arises if the  metric has             
 a conformal killing vector, i.e. if the  metric is left invariant upto a weyl  transformation
by the corresponding   coordinate transformation. In the solutions of  this section and in \S4           
and \S5, we construct examples where the metric has genuine killing vectors  which make it  
 homogenoeus along the field theory directions while possessing an  additional  
 conformal killing vector which leads  to a scaling transformation  in the field theory.

\subsection{General idea}{\label{subsec:generalidea}}
In this subsection we discuss the general prescription of obtaining hyperscaling violating solutions from scaling solutions
via dimensional reduction, as described in \cite{Gouteraux:2011qh}.
Let us consider the metric ansatz eq.(\ref{ansatzg}), where $\alpha_1,\alpha_2$ are constants given by eq.(\ref{alpha_2alpha_1byd}). 
Let $r$ be the ``radial" coordinate in $(d+1)$-dimensions,
which will also appear as a coordinate after compactification to $d$-dimensions. We will consider metric components to be functions of $r$ only, and 
will allow the 
higher dimensional metric to be invariant under scale transformations, generated by a shift of $r$ (and appropriate rescaling of the spatial coordinates). 
Let $\phi(r)=\phi_0 r$:
\begin{equation}
 d \hat s^2 = e^{2 \alpha_1 \phi_0 r} ds^2+ e^{2 \alpha_2 \phi_0 r} \left(dz+B_{\mu}(r) dx^{\mu}\right)^2
\end{equation}
Now, requiring that $d \hat s^2$ is invariant under $r \to r+\epsilon$ implies that the lower dimensional metric transform as
\begin{equation}
 ds^2 = e^{-2 \alpha_1 \phi_0 \epsilon} ds^2
\end{equation}
So the lower dimensional metric is invariant under the scale transformation up to an overall scaling function in front of the metric, a conformal factor.
This corresponds to a metric which exhibits hyperscaling violation.

\subsection{Dimensional reduction of 5d type III to 4d space-time}{\label{subsec:dimred}}

In this subsection, we consider the dimensional reduction of the type III metric as given in
eq.(4.38) of \cite{Bianchi}. 
Let us first recall the relevant results of \cite{Bianchi}. 

The action is given by eq.(\ref{EHaction_V_L}) with $d=4$ and only one  vector field turned on.
For convenience we do not write the quantities with $\hat ~$ here, and we explicitly mention the space-time
dimension of the quantity. The metric in $(4+1)$ dimensions takes the form
\begin{equation}
ds^2=dr^2-e^{2\beta_t r} dt^2 + e^{2\beta_2 r} dz^2 + {1\over \rho^2}
(d\rho^2+ dx^2)
\end{equation}
with vector field  given by
\begin{equation}
 A = \sqrt{A_t} e^{\beta_t r} dt ~.
\end{equation}
In order to find a solution, the parameters in the action and in the metric/vector field should be related via:
\begin{eqnarray}
m^2&=&2(1-\beta_2^2) \quad \,, \quad 
\Lambda = {1\over \beta_2^2}+2\beta_2^2 \,, \\
A_t & = & {2-4\beta_2^2\over 1-\beta_2^2} \quad \,, \quad 
\beta_t  =  {1-\beta_2^2\over \beta_2} \,.
\end{eqnarray}

Now, consider the dimensional reduction along $z$ by using the techniques of \S {\ref{subsec:generalidea}}. 
For this subsection we use $\alpha_1 = { 1 \over 2 \sqrt{3}}$, $\alpha_2 = -{1 \over \sqrt{3}}$ from eq.(\ref{alpha_2alpha_1byd}) with $d=4$.

The new 4d action is given by eq.(\ref{dimredaction})
\begin{equation}\label{4dactiontypeiii}
  S = \int d^4 x \sqrt{-g} \left\lbrace R - {1 \over 2} (\nabla \phi)^2+ e^{2
\alpha_1
\phi} \Lambda - {1 \over 4} e^{-2 \alpha_1 \phi}F^2 - {1 \over 4} m^2  A^2
\right\rbrace~.
\end{equation}
The relevant solution can be read off from the 5d solution above. The metric is 
\begin{equation}
\label{thisistypeIIIin4d}
 ds^2 = e^{\beta_2 r} \left( dr^2 - e^{2 \beta_t r} dt^2 + {d\rho^2 + dx^2 
\over \rho^2} \right) \,, 
\end{equation}
and the matter fields are
\begin{equation}
 A = \sqrt{A_t} e^{ \beta_t r} dt \quad \,, \quad \phi = {\beta_2 \over \alpha_2} r ~.
\end{equation}
Actually eq.(\ref{thisistypeIIIin4d}) shows that this is a hyperscaling violating solution 
of Bianchi type III.

It is possible to view this solution in different coordinates which make it more
transparent that it is a hyperscaling violating solution. 
We perform the following coordinate transformation
\begin{equation}
 r  = {1 \over \beta_2+ \beta_t} \log{\tilde{r}} \,.
\end{equation}
In the new coordinates, the solution is given by
\begin{eqnarray}\label{type34deqns}
 ds^2 &=& \left({1 \over \beta_2+ \beta_t}\right)^2{d\tilde{r}^2 \over \tilde{r}^{2
\gamma}} - \tilde{r}^{2 \gamma} dt^2 + \tilde{r}^{2\beta}  \ {d\rho^2 + dx^2 
\over \rho^2} \\
A &=& \sqrt{A_t} \ \tilde{r}^{\beta_t \over \beta_2 + \beta_t} \ dt \,, \\
\phi &=& {\beta_2 \over \alpha_2 (\beta_2+ \beta_t)} \ \log \tilde r \,,
\end{eqnarray}
where 
\bea
2 \gamma ={\beta_2+ 2 \beta_t \over \beta_2+ \beta_t} \quad \,, \quad 2 \beta = {\beta_2
\over \beta_2+ \beta_t} \,.
\eea

For general mass parameter, we have $\beta_2 \neq 0$. Therefore, we generically have 
$\gamma \neq 1$, which implies that this metric is not scale invariant. Instead, it
Weyl rescales under scale transformations, in the way which is characteristic of hyperscaling violating metrics.

%%%%%%%%%%%%%%%%%%%%%%%%%%%%%%%%%%%%%%%%%%%%%%%%%%%%

\subsection{Hyperscaling violating Type III metrics in 4d space-time}{\label{subsec:typeIII}}

We saw  that the Kaluza-Klein reduced action
eq.(\ref{4dactiontypeiii}) has a hyperscaling violating type III solution of the
form eq.(\ref{type34deqns}). In this subsection, we define a family of actions which
generalizes the 4d action eq.(\ref{4dactiontypeiii}) of the previous subsection, and we
exhibit more generic hyperscaling violating type III solutions of these actions.   We note that generic
values of the parameters here do not allow an ``uplift" to a 5d solution, so
these solutions considerably generalize those of \S \ref{subsec:dimred}.

Consider the general Einstein Hilbert action in $(d+1)$-dimensions coupled to
a massless scalar field, a massive vector field and a cosmological constant:
\begin{equation}~\label{EHAction_S_V_L}
 S = \int d^{d+1} x \sqrt{g} \left\lbrace R - {1 \over 2} (\nabla \phi)^2  -
{e^{2 \alpha \phi} \over 4} F^2 - { m^2 \over 4} e^{2 \epsilon \phi}  A^2 + e^{2
\delta \phi}
\Lambda \right\rbrace~.
\end{equation} 
In this section we will consider $d=3$. This action is a natural generalization of the one given by eq.(\ref{4dactiontypeiii}). 

As an ansatz, let us take  a hyperscaling violating type III metric of the form:
\begin{equation}
 ds^2 = \lambda^2 {dr^2 \over r^{2 \gamma}} -  r^{2 \gamma} dt^2 + {r^{2 \beta} \over \rho^2}
(d\rho^2 + dx^2) \,.
\end{equation}
Also, let the gauge field and the scalar field be 
\begin{equation}
 A = \sqrt{A_t}   r^\theta dt \,, \quad 
 \phi = k \log(r)~.
\end{equation}

All of the equations become algebraic if we take the following relations among parameters:
\begin{equation}
 \gamma =  1 + k \delta \quad \,, \quad \epsilon=\alpha +\delta \quad \,, \quad \theta =
1 + k (\delta-\alpha) \quad \,, \quad \beta = - k \delta ~.
\end{equation} 
Plugging in these relations, the Einstein equations 
along $r,t,\rho$ become   
\begin{eqnarray}
 -8 k \delta  (2+k \delta )+A_t \left(-m^2 \lambda^2 +2 (1+k (-\alpha +\delta ))^2\right)-2 \left(-4 \lambda^2+k^2+2 \Lambda \lambda^2\right) &=& 0 \,, \quad \quad \\
 A_t \left(m^2 \lambda^2+2 (1-k \alpha + k \delta )^2\right)+2 \left(4 \lambda^2 +k^2 \left(1+4 \delta ^2\right)-2 \Lambda \lambda^2 \right) &=& 0 \,, \quad \quad \\
 8 k \delta  (2+ k \delta )-A_t\left(m^2 \lambda^2+2 (1+k (-\alpha +\delta ))^2\right)
 +2 \left(4   +k^2-2 \Lambda \lambda^2 \right) &=& 0  \,. \quad \quad 
\end{eqnarray}
The equation along $x$ is the same as the equation along $\rho$ because $(\rho,x)$ span an Euclidean $AdS_2$ factor.  

The gauge field equation is
\begin{equation}
m^2 \lambda^2 +2 k (-1+k (\alpha -\delta )) (\alpha -\delta )=0\,,
\end{equation}
and the scalar field equation is 
\begin{equation}
2 k+ A_t \left(2 \alpha +2 k^2 \alpha  (\alpha -\delta )^2+4 k \alpha  (-\alpha +\delta )+m^2 \lambda^2(\alpha +\delta )\right)+4 \delta  \lambda^2 \Lambda = 0~.
\end{equation}

The solutions to the above equations are
\begin{eqnarray}
&&\epsilon = \alpha+ \delta \,, \hspace{10mm} 
\gamma = \frac{1+2 (\alpha -\delta ) \delta  \left(1+\lambda ^2\right)}{1+4 \alpha  \delta } \,,  \hspace{10mm} 
\theta = 1+\frac{2 \left(\alpha ^2-\delta ^2-(\alpha -\delta )^2 \lambda ^2\right)}{1+4 \alpha  \delta } \,, \hspace{10mm}  \nonumber \\
&& \beta = \frac{2 \delta  \left(\alpha +\delta -\alpha  \lambda ^2+\delta  \lambda ^2\right)}{1+4 \alpha  \delta } \,, \hspace{10mm}     
k =  \frac{-2 (\alpha +\delta )+2 (\alpha -\delta ) \lambda ^2}{1+4 \alpha  \delta } \,, \nonumber \\
&& A_t =\frac{2 \left(-1+\lambda ^2+4 \delta ^2 \left(1+\lambda ^2\right)\right)}{-1-2 \alpha ^2-4 \alpha  \delta +2 \delta ^2+2 (\alpha -\delta )^2 \lambda ^2} \,, \nonumber \\
&& \Lambda =  \frac{1}{{(\lambda +4 \alpha  \delta  \lambda )^2} } 
\times  \Bigl\{    {\left(1+\lambda ^2\right) \left(1+2 \delta ^2 \left(-1+\lambda ^2\right)+8 \delta ^4 \left(1+\lambda ^2\right)\right) 
 }     \nonumber \\
&& \quad  \quad \quad {   
+2 \alpha ^2 \left(1-2 \lambda ^2+\lambda ^4+4 \delta ^2 \left(1+\lambda ^2\right)^2\right)    
 -4 \alpha  \delta  \left(-2+\lambda ^2 \left(-3+\lambda ^2+4 \delta ^2 \left(1+\lambda ^2\right)\right)\right)}  \Bigr\} \,,
\nonumber \\
&& m^2 =-\frac{4 (\alpha -\delta ) \left(-1-2 \alpha ^2-4 \alpha  \delta +2 \delta ^2+2 (\alpha -\delta )^2 \lambda ^2\right) \left(\alpha  \left(-1+\lambda ^2\right)-\delta  \left(1+\lambda ^2\right)\right)}{(\lambda +4 \alpha  \delta  \lambda )^2}~ . \quad
\end{eqnarray}

Throughout this paper we will take  the cosmological constant to be negative,  
this corresponds to taking $\Lambda >0$ in our conventions\footnote{Hopefully this will allow  
the near-horizon geometries we study to smoothly connect  with asymptotic AdS space. We leave a 
study of such   interpolating geometries for the future.}. In addition we will mainly consider
geometries where the horizon area vanishes\footnote{Of course, as the extremal RN black brane example shows, this
 condition can be relaxed in some cases.}. 
Choosing the horizon to lie at $r \rightarrow 0$ this gives rise to the conditions
$\gamma, \beta>0$. Finally, for a physically acceptable solution $A_t, \lambda>0$. 
All these conditions can be met for the solution found above in various open sets of parameter ranges;
one such range is
$\alpha <\frac{1}{2} \left(-8-3 \sqrt{7}\right)$, $-\frac{1}{2}<\delta <0$, $~0<\lambda <\sqrt{\frac{1-4 \delta ^2}{1+4 \delta ^2}}$.

%For the solution to be ``physically acceptable''\footnote{\label{PhysicallyAcceptable}Let us define the notion of ``physically acceptable'' solutions which we use again and again throughout the paper. 
%In all the solutions we always require $\Lambda$ (=minus of the cosmological constant) to be positive so that it smoothly connect to asymptotic AdS. 
%We always demand that the various parameters of the solutions to be real, and  
%also the existence of a horizon, by ensuring that the time component of
%the metric $g_{tt}$ vanishes at the horizon. Finally, we also demand
%that the volume of the hypersurface at constant time and radial coordinate (the ``field theory" spatial slices) should decrease towards the horizon. This ensures that the entropy vanishes
%at the horizon, implying the Nernst's law in boundary theory viewpoint.}, 
%the conditions are $\Lambda>0$, $\beta>0$, $\gamma>0$, $A_t > 0$, $\lambda>0$. This allows for various open sets of parameter ranges;
%one such range is 
%$\alpha <\frac{1}{2} \left(-8-3 \sqrt{7}\right)$, $~-\frac{1}{2}<\delta <0$, $~0<\lambda <\sqrt{\frac{1-4 \delta ^2}{1+4 \delta ^2}}$.

\section{Hyperscaling violation in 5d Bianchi VI, III, V attractors}\label{sec:hyperVI}

In \S \ref{sec:hypervioKK}, we have seen how dimensional reduction of the Bianchi solutions of
\cite{Bianchi} can give rise to hyperscaling violating
solutions. This leads to an expectation that such solutions
are fairly easy to find in their own right, also in 5d.  We confirm this expectation in
this section by finding hyperscaling violating generalizations for many of the solutions found in
\cite{Bianchi}.  In this section, we exhibit hyperscaling violating solutions in 5d for
Bianchi types VI, III and V. 
We leave the discussion of type VII$_0$, which has been especially popular
in the literature, to a separate section, \S \ref{sec:hyperVII}.

We do the calculations in 5d, although we expect similar results to hold in other
dimensions too. We begin with the action given by eq.(\ref{EHAction_S_V_L}) for $d=4$.
We can achieve hyperscaling violation by allowing the scalar field to run logarithmically with the radial coordinate. 
This is similar to the earlier dilatonic realizations of hyperscaling violating geometries studied in \cite{lifsol3, hv1, hv2}.

Before we move on to individual solutions, it is worth noting that the  
dimensionless constants in the action are $\alpha,\delta, \epsilon$ and ${m^2 \over \Lambda}$. In the discussion that follows
we will  find it convenient to take   $\epsilon=\alpha+\delta$ since the equations simplify in this case. 
%(this is because we will see later that $\epsilon$ is constrained as $\epsilon=\alpha+\delta$) 
%and the dimensionless coupling
%constant $m^2 \over \Lambda$. 
Furthermore, sometimes the algebra governing the
Bianchi type will have some free
parameters too -- for example, type VI has one parameter $h$ appearing in its real three-algebra.

%%%%%%%%%%%%%%%%%%%%%%%%%%%%%%%%%%%%%%%%%%%
\subsection{Type VI (generic $h$)}
To begin with, let us start with a Type VI metric. We take the ansatz for the metric and matter fields to be
of the form
\begin{eqnarray}
 ds^2&=& {dr^2 \over r^{2 \gamma} } -r^{2 \gamma} dt^2+ \lambda r^{2 \beta_x} dx^2 +
r^{2 \beta_y} e^{-2 x}dy^2 + r^{2 \beta_z} e^{-2 h x}dz^2 \,,\\
A &=& \sqrt{A_t}  r^\theta dt  \,, \\
\phi &=& k \log(r) \,.
\end{eqnarray}

All  of the equations become algebraic if we choose
\begin{eqnarray}
\beta_x =  -\delta  k \quad \,, \quad \gamma =  1+ \delta  k \quad \,, \quad \epsilon
= \text{ }\alpha +\delta \quad \,, \quad \theta = 1+ k (\delta -\alpha )~.
\end{eqnarray}

%The scalar field equation is 
%\begin{equation}
%{1 \over \sqrt{g}}{ \partial_\mu (\sqrt{g} \partial^\mu \phi ) } - {\alpha \over 2} e^{2
%\alpha \phi} F^2 - {\epsilon \over 2}  m^2 e^{2 \epsilon \phi} A^2 + 2 \delta
%\Lambda e^{2 \delta \phi} = 0~.
%\end{equation}

%The gauge field equation is
%\begin{equation}
%{1 \over \sqrt{g}} { \partial_\mu (\sqrt{g}e^{2
%\alpha \phi}  F^{\mu \nu}) }  - { m^2 \over 2} e^{2 \epsilon \phi}
%A^\nu  = 0~.
%\end{equation}

The Einstein equations, gauge field equation, and scalar field equation then respectively give:
\begin{footnotesize}
\begin{eqnarray}\label{type6eqns1}
\nonumber
 A_t (2 (1+k \delta-k \alpha)^2-m^2)+8 (\beta_y +\beta_z+\beta_y
\beta_z)  
-2 k \left(k+4 \delta +4 k \delta ^2\right)+\frac{8
\left(1+h+h^2\right)}{\lambda }-4 \Lambda  &=& 0 \,,    
\nonumber \\
\nonumber
A_t (-2 (1+k \delta-k \alpha)^2-m^2)-\frac{8 \left(1+h+h^2\right)}{\lambda }-2 \left(k^2+4
\left(\beta_y^2+\beta_y \beta_z+\beta_z^2\right)-2
\Lambda \right) &=& 0  \,,   
\\
\nonumber
A_t (-2 (1+k \delta-k \alpha)^2-m^2)+\frac{8 h}{\lambda } \hspace{88mm} && \nonumber \\
+2 \left(4 \left(1+\beta_y+\beta_y^2+\beta_z+\beta_y
\beta_z+\beta_z^2\right)+k \left(k+4 (3+2 \beta_y+2 \beta_z) \delta
+8 k \delta ^2\right)-2 \Lambda \right) &=& 0 \,, \nonumber \\
A_t (-2 (1+k \delta-k \alpha)^2-m^2)   
    +\frac{8 h^2}{\lambda }+2 \left(4+4 \beta_z
(1+\beta_z)+k \left(k+4 (2+\beta_z) \delta +4 k \delta ^2\right)-2
\Lambda \right) &=& 0 \,, \\
\nonumber
A_t (-2 (1+k \delta-k \alpha)^2-m^2)   
 +\frac{8}{\lambda }+2 \left(4+4 \beta_y
(1+\beta_y)+k \left(k+4 (2+\beta_y) \delta +4 k \delta ^2\right)-2
\Lambda \right)&=& 0  \,, \nonumber \\
\nonumber
 \beta_y+h \beta_z+(1+h) k \delta  &=& 0 \,, \nonumber \\
\nonumber
 m^2-2 (k \alpha +\beta_y+\beta_z) (1+ k \delta-k \alpha )
&=& 0 \,, \\
\nonumber
\label{type6eqns8}
 2 k (1+\beta_y+\beta_z+k \delta )+A_t \left[ m^2( \alpha +\delta)
+2 \alpha (1+ k \delta - k \alpha)^2 \right]+4 \delta  \Lambda &=& 0 \,.
\end{eqnarray}
\end{footnotesize}

Generically, it is  difficult to find solutions to these nonlinear algebraic equations. 
However fortunately, given a choice of $h$ which characterizes the three-algebra, 
one can show that the following solutions exist:
\begin{footnotesize}
\begin{eqnarray}
\nonumber
\label{generichsolutionfor5dbianchiIII}
\beta_z &=& \frac{k (2 \delta  (-\alpha  (h (h+2)-1)+\delta  h (2 h+1)+\delta )-h+1)-2
   (h-1) (\alpha +\delta )}{2 \left(h^2+1\right) (\alpha -2 \delta )} \,, \\
\nonumber
\beta_y &=& \frac{k \left(2 \alpha  \delta  ((h-2) h-1)+2 \delta ^2
   \left(h^2+h+2\right)+(h-1) h\right)+2 (h-1) h (\alpha +\delta )}{2
   \left(h^2+1\right) (\alpha -2 \delta )}  \,, \\
\nonumber
\lambda &= & \frac{4 \left(h^2+1\right)^2 (\alpha -2 \delta )^2}{\left(\alpha  (4 \delta 
   k+2)-2 \delta ^2 k+2 \delta +k\right)} \nonumber \\
   && \times \frac{1}{ \left(k \left(2 \alpha  \delta 
   ((h-4) h+1)+2 \delta ^2 (h+1)^2+(h-1)^2\right)+4 \alpha  ((h-1) h+1)-2
   \delta  (h+1)^2\right)}  \,,  \\
\nonumber
A_t &=& \frac{ \left(k+6 \delta +6 k \delta^2\right)}{(-1+k (\alpha -\delta )) (\alpha -2 \delta) } \,, \\
\Lambda & =& \frac{1}{4 \left(h^2+1\right)  (\alpha -2   \delta )^2} 
\nonumber \\
&& \times
   \left[ {4 k \{  6 \alpha ^2 \delta  (h (2 h-3)+2)-\alpha  \left(3 \delta ^2
   (h (h+6)+1)-4 h^2+6 h-4\right)+\delta  \left(12 \delta ^2
   \left(h^2+1\right)+(h-6) h+1\right) \}
   } \right. \nonumber \\
   &&\left. { 
   +k^2 \{
   2 \alpha ^2
   \left(12 \delta ^2 (h-1)^2+h^2+1\right)
   -6 \alpha  \delta  \left(2 \delta^2 (h+1)^2-(h-6) h-1\right)  
   } \right.
   \nonumber \\ 
   && 
    \left.   { +12 \delta ^4 (h (3 h+2)+3)+16 \delta ^2
   \left(h^2+1\right)   +3 (h-1)^2
   \}        
   +12 \{  2 \alpha ^2 ((h-1)
   h+1)-4 \alpha  \delta  h+\delta ^2 (h-1)^2 \} 
   } \right] \,, \quad 
   \nonumber \\
\nonumber
 m^2 &=& (1+\delta  k-\alpha k) ~ \frac{k \left(2 \alpha ^2 \left(h^2+1\right)-4 \alpha  \delta 
   (h+1)^2+\delta ^2 \left(6 h^2+4 h+6\right)+(h-1)^2\right)+2 (h-1)^2
   (\alpha +\delta )}{\left(h^2+1\right) (\alpha -2 \delta ) } \,.
\end{eqnarray}
\end{footnotesize}
Note that $A_t$ is $h$ independent, but the rest of the formulae do depend on $h$. 

This messy formula gives, for example, with $h = {1\over 2}$:
\begin{eqnarray}
\nonumber
 \beta_z &=&\frac{k+2 \alpha +2 \delta -k \alpha  \delta +8 k \delta ^2}{5 \alpha -10
\delta }  \,, \\
\nonumber
\beta_y &=& \frac{k+2 \alpha +2 \delta +14 k \alpha  \delta -22 k \delta ^2}{10
(-\alpha +2 \delta )}  \,, \\
\nonumber
\lambda &=& -\frac{25 (\alpha -2 \delta )^2}{(-12 \alpha +18 \delta +k (-1+6 (\alpha
-3 \delta ) \delta )) \left(2 (\alpha +\delta )+k \left(1+4 \alpha  \delta -2 \delta
^2\right)\right)} \,, \quad \quad \\
A_t &=& \frac{ \left(k+6 \delta +6 k \delta^2\right)}{(-1+k (\alpha -\delta )) (\alpha -2 \delta) } \,, \\
\nonumber
\Lambda &=& \frac{1}{20 (-\alpha +2 \delta )^2} \nonumber \\
& &  \times \{  {12 \left(6 \alpha ^2-8 \alpha  \delta +\delta ^2\right)+4 k \left(8
\alpha -7 \delta +24 \alpha ^2 \delta -51 \alpha  \delta ^2+60 \delta ^3\right) } \nonumber \\
&& \quad \quad { +k^2
\left(3+80 \delta ^2+228 \delta ^4+2 \alpha ^2 \left(5+12 \delta ^2\right)-6 \alpha 
\delta  \left(7+18 \delta ^2\right)\right)} \} \,, \nonumber \\
\nonumber
m^2 &=& (1-k (\alpha -\delta ))~ \frac{2 (\alpha +\delta )+k \left(1+10 \alpha ^2-36 \alpha  \delta +38 \delta
^2\right)}{5  (\alpha -2 \delta )} \,.
\end{eqnarray}
We take $\Lambda>0$ and also $\gamma>0$. An acceptable solution with vanishing horizon area then requires
$(\beta_x+\beta_y+\beta_z)$, $\lambda$, $A_t >0$. 
One  possible range  where these conditions is met is 
$0<\alpha \leq \frac{\sqrt{6}}{5}$, $0<\delta < \frac{\alpha}{2}$, $-\frac{2 (\alpha +\delta )}{1+4 \alpha  \delta -2 \delta^2}<k<-\frac{6 \delta }{1+6 \delta^2}$.

%One can also see that these solutions are physically acceptable\footnoteref{PhysicallyAcceptable} by verifying that
%$(\beta_x+\beta_y+\beta_z)$, $\lambda$, $A_t$, $\Lambda$, $\gamma$ $> 0$. One such possible range is
%$0<\alpha \leq \frac{\sqrt{6}}{5}$, $0<\delta < \frac{\alpha}{2}$, $-\frac{2 (\alpha +\delta )}{1+4 \alpha  \delta -2 \delta^2}<k<-\frac{6 \delta }{1+6 \delta^2}$.

%%%%%%%%%%%%%%%%%%%%%%%%%%%%%%%%%%%%%%%%%%%
\subsection{Type III ($h=0$)}

This is a limiting case of type VI, with $h=0$. The solution 
is obtained from eq.(\ref{generichsolutionfor5dbianchiIII})  by setting $h=0$:
\begin{eqnarray}
\nonumber
 \beta_z &=& \frac{2 (\alpha +\delta )+k (1+2 \delta  (\alpha +\delta ))}{2
(\alpha -2 \delta )}  \,, \\
\nonumber
\beta_y &=& - k \delta \,, \\
\nonumber
\lambda &=& \frac{4 (\alpha -2 \delta )^2}{\left(2 (\alpha +\delta )+k \left(1+4
\alpha  \delta -2 \delta ^2\right)\right) (4 \alpha -2 \delta +k (1+2 \delta  (\alpha
+\delta )))} \,, \\
A_t &=& \frac{ \left(k+6 \delta +6 k \delta^2\right)}{(-1+k (\alpha -\delta )) (\alpha -2 \delta) }  \,, \\
\nonumber
\Lambda &=& \frac{1}{4 (\alpha -2 \delta )^2}  \nonumber \\
&& \times \{12 \left(2 \alpha ^2+\delta ^2\right)+k^2 \left(3+2 \alpha ^2+6
\alpha \delta +8 \left(2+3 \alpha ^2\right) \delta ^2-12 \alpha  \delta ^3+36 \delta
^4\right) \nonumber \\
&&\quad \quad +4 k \left(\delta +12 \alpha ^2 \delta +12 \delta ^3+\alpha  \left(4-3 \delta
^2\right)\right) \} \,,
\nonumber \\
m^2 &=& (1-k (\alpha -\delta )) ~ \frac{2 (\alpha +\delta )+k \left(1+2 \alpha ^2-4 \alpha  \delta +6
\delta ^2\right)}{(\alpha -2 \delta )} \,. \nonumber 
\end{eqnarray}

\noindent
An acceptable solution with vanishing horizon area arises if 
$(\beta_x+\beta_y+\beta_z)$, $\lambda$, $A_t$, $\Lambda$, $\gamma > 0$. 
These conditions can be met, e.g., one possible range of parameters is
$0<\alpha <\sqrt{\frac{2}{3}}$, $0<\delta <\frac{\alpha }{2}$, $\frac{-2 \alpha -2 \delta }{1+4 \alpha  \delta -2 \delta^2}<k<-\frac{6 \delta }{1+6 \delta^2}$.

%%%%%%%%%%%%%%%%%%%%%%%%%%%%%%%%%%%%%%%%%%%

\subsection{Type V ($h=1$)}
This is a limiting case of type VI, with $h=1$. The solution 
is obtained from eq.(\ref{generichsolutionfor5dbianchiIII})  by setting $h=1$:
\begin{eqnarray}
\beta_y &=&  \beta_z = - k \delta \,, \nonumber \\
\lambda &=& -\frac{4 (\alpha -2 \delta )}{(-1+k \delta ) \left(2 (\alpha +\delta )+k
\left(1+4 \alpha  \delta -2 \delta ^2\right)\right)} \,, \nonumber \\
A_t &=& \frac{ \left(k+6 \delta +6 k \delta^2\right)}{(-1+k (\alpha -\delta )) (\alpha -2 \delta) }  \,, \\
\Lambda &=&  \frac{6 \alpha +k \left(2+6 (\alpha -2 \delta ) \delta +k \left(\alpha -4
\left(\delta +3 \delta ^3\right)\right)\right)}{2 (\alpha -2 \delta )}  \,, \nonumber \\
m^2 &=& 2 k (\alpha -2 \delta ) (1+k (-\alpha +\delta )) \,. \nonumber
\end{eqnarray}

\noindent
Again, an acceptable solution with vanishing horizon area arises when
 $(\beta_x+\beta_y+\beta_z)$, $\lambda$, $A_t$, $\Lambda$, $\gamma> 0$.  We can find
some allowable range of values for $\alpha$, $\delta, k$ where these conditions are met. For example: $0<\alpha <\sqrt{\frac{2}{3}}$, $0<\delta <\frac{\alpha }{2}$, $\frac{-2 \alpha -2 \delta }{1+4 \alpha  \delta -2 \delta^2}<k<-\frac{6 \delta }{1+6 \delta^2}$.

%%%%%%%%%%%%%%%%%%%%%%%%%%%%%%%%%%%%%%%%%%%%%%%%%%%%%%

\section{Hyperscaling violation in 5d Bianchi VII$_0$ attractors}\label{sec:hyperVII}

In this section we will obtain a hyperscaling violating solutions in 5d with
Bianchi type VII$_0$ symmetry using the same theory as before, i.e eq.(\ref{EHAction_S_V_L}) with $d=4$. Again we
will have $\epsilon = \alpha + \delta$, so that the solutions will have $\alpha,
\delta, {m^2 \over \Lambda}$ as free parameters. The ansatz for the various fields is chosen as
\begin{eqnarray}
ds^2 &=&  C_a^2{ dr^2 \over r^{2 \gamma} } -r^{2 \gamma} dt^2+ r^{2 \beta_x} dx^2 +
r^{2 \beta_{yz}} \left(\left(\omega^2\right)^2+\lambda
\left(\omega^3\right)^2\right)~. \\
 A &=& \sqrt{A_2} \ r^\theta \omega^2. \\
 \phi &=& k \log(r)~.
\end{eqnarray}
Here, the $\omega^i$ are two of the invariant one-forms of Type VII$_0$ geometry:
\begin{equation}
 \omega^2 = \cos(x) dy + \sin(x) dz   \hspace{20mm} \omega^3 =-\sin(x) dy + \cos(x)
dz~.
\end{equation}

All the equations of motion become algebraic  if we choose
\begin{equation}
 \beta_x = -k \delta \,, \hspace{5mm} \gamma=1+ k \delta \,, \hspace{5mm} 
 \epsilon=\alpha+\delta \,, \hspace{5mm} \theta=\beta_{yz}-k \alpha \,.
\end{equation}
The Einstein equations, gauge field, and scalar equations respectively give:
\begin{footnotesize}
\begin{eqnarray} 
\nonumber
 -2 \left(\beta_{yz} (-8+(-4+A_2) \beta_{yz})+k (-2 A_2
\alpha  \beta_{yz}+4 \delta )+k^2 \left(1+A_2 \alpha ^2+4 \delta
^2\right)\right) \lambda \hspace{15mm}
&& \nonumber \\ %%
+C_a^2 (2+A_2 (2+{m^2} \lambda )+2 \lambda 
(-2+\lambda -2 \Lambda ))  &=&0 \,, \nonumber \\
\nonumber
2 \left(k^2 \left(1+A_2 \alpha ^2\right)-2 A_2 k \alpha  
\beta_{yz} 
+(12+A_2) \beta_{yz}^2\right) \lambda \hspace{55mm}
&& \nonumber \\ %%
+C_a^2 (2+A_2
(2+{m^2} \lambda )+2 \lambda  (-2+\lambda -2 \Lambda ))  &=&0 \,, \nonumber \\
\nonumber
2 \left(4+\beta_{yz} (8+(12+A_2) \beta_{yz})+2 k (-A_2
\alpha  \beta_{yz}+6 \delta +8 \beta_{yz} \delta )+k^2 \left(1+A_2
\alpha ^2+8 \delta ^2\right)\right) \lambda 
&& \nonumber \\ %%
+C_a^2 (-2+A_2 (-2+{m^2}
\lambda )-2 \lambda  (-2+\lambda +2 \Lambda )) &=&0 \,,   \\
\nonumber
 2 \left(4+2 A_2 k \alpha  \beta_{yz}+
  \beta_{yz} 
 (4+4 \beta_{yz}-A_2 \beta_{yz})+4 k (2+\beta_{yz})
\delta +k^2 \left(1-A_2 \alpha ^2+4 \delta ^2\right)\right) \lambda \hspace{1mm}
&& \nonumber \\ %%
-C_a^2
(A_2 (-2+{m^2} \lambda )+2 (-3+\lambda  (2+\lambda +2 \Lambda ))) 
&=& 0 \,, \nonumber \\
\nonumber
2 \left(4-2 A_2 k \alpha  \beta_{yz}+\beta_{yz} (4+(4+A_2)
\beta_{yz})+4 k (2+\beta_{yz}) \delta +k^2 \left(1+A_2 \alpha ^2+4
\delta ^2\right)\right) \lambda \hspace{4mm}
&& \nonumber \\ %%
+C_a^2 \left(-2+6 \lambda ^2+A_2
(-2+{m^2} \lambda )-4 \lambda  (1+\Lambda )\right) &=& 0 \,, \nonumber \\
\nonumber
\lambda  (2 (k \alpha -\beta_{yz}) (1+\beta_{yz}+k (\alpha +\delta )) \lambda 
+C_a^2 (2+m^2 \lambda ))&=&0  \,, \\
\nonumber
 -A_2 \left(2 C_a^2 \alpha +2 \alpha  (-k \alpha +\beta_{yz})^2 \lambda +C_a^2 m^2 (\alpha +\delta ) \lambda \right)+2 \lambda  \left(k+2 k \beta_{yz}+k^2 \delta +2 C_a^2 \delta  \Lambda \right)&=&0~.
\end{eqnarray}
\end{footnotesize}
Note that all of the 7 equations are written in vierbein coordinates. Six of the
equations are independent, and can be used to solve for the six
parameters $C_a^2$, $\Lambda$, $\beta_{yz}$, $k$, $m^2$, $A_2$ in terms of $\alpha$, $\delta$,
$\lambda$. 
Although we were able to solve above equations in terms of $\alpha$, $\delta$,
$\lambda$, the solutions are very complicated and  not very illuminating.
Instead we  present below  solutions at special values in the $(\alpha, \delta)$ parameter 
space, where the resulting expressions are more compact. 

\medskip

\noindent
{\bf Solution for $\alpha=-\delta$:}
 {\footnotesize
 \begin{eqnarray}
\nonumber
C_a^2&=& \frac{2 \left(-1+6 \delta ^2\right) \left(11+12 \delta ^2 (-1+\lambda )-5 \lambda \right) (-2+\lambda )}{\left(1+6 \delta ^2\right)^2 (-3+\lambda )^2 (-1+\lambda )} \,, \\
\nonumber
  \beta_{yz}&=& \frac{2 \left(-2-3 \delta ^2 (-1+\lambda )+\lambda \right)}{\left(1+6 \delta ^2\right) (-3+\lambda )} \,, \\
  k&=& -\frac{6 \delta }{1+6 \delta ^2} \,, \\
  m^2 &=& \frac{-22+24 \delta ^2 (-1+\lambda )^2-2 \lambda  (2+\lambda  (-10+3 \lambda ))}{\left(11+12 \delta ^2 (-1+\lambda )-5 \lambda \right) \lambda } \,, \nonumber \\
\nonumber
  \Lambda &=& \frac{(-1+\lambda ) \left(-22+144 \delta ^4 (-1+\lambda )^3+\lambda  (117+\lambda  (-90+19 \lambda ))+12 \delta ^2 (13+\lambda  (-24+(24-7 \lambda ) \lambda ))\right)}{2 \left(-1+6 \delta ^2\right) \left(11+12 \delta ^2 (-1+\lambda )-5 \lambda \right) (-2+\lambda ) \lambda } \,, \\
\nonumber
  A_2&=& -2-\frac{2}{-2+\lambda }~.
 \end{eqnarray}
}

A solution with vanishing horizon area would arise if 
 $C_a^2>0$, $A_2>0$, $\beta_x+2 \beta_{yz}>0$, $\Lambda>0$, $\lambda>0$ are satisfied. 
These conditions are indeed met when  $-\frac{1}{\sqrt{6}}<\delta <\frac{1}{\sqrt{6}}$, $1<\lambda <2$.
This specific case $\alpha=-\delta$ can be obtained from dimensional reduction, but we can get more generic solutions, which do
not allow an uplift to higher dimensions, as follows.
\\ \\

\begin{itemize}
\item {\bf Solution for $\alpha=\delta=-1$:}
{\scriptsize
 \begin{eqnarray}\nonumber
  C_a^2 &=& \frac{-81 \left(-12+\sqrt{144+\lambda  (-112+\lambda  (113+9 \lambda  (-14+9 \lambda )))}\right)}{49 (-1+\lambda ) (27+\lambda  (-14+23 \lambda ))^2} 
 \nonumber \\ 
 & +& \frac{\lambda  \left(2793-938 \sqrt{144+\lambda  (-112+\lambda  (113+9 \lambda  (-14+9 \lambda )))}\right)}{49 (-1+\lambda ) (27+\lambda  (-14+23 \lambda ))^2} \nonumber \\
&+& \frac{\lambda^2  \left(-11965+\lambda  (-11109+9589 \lambda )+1991 \sqrt{144+\lambda  (-112+\lambda  (113+9 \lambda  (-14+9 \lambda )))}\right)}{49 (-1+\lambda ) (27+\lambda  (-14+23 \lambda ))^2}  \,,
\nonumber \\
\beta_{yz}&=& \frac{30+\lambda  (-21+37 \lambda )+\sqrt{144+\lambda  (-112+\lambda  (113+9 \lambda  (-14+9 \lambda )))}}{7 (27+\lambda  (-14+23 \lambda ))} \,, \nonumber \\
k &=& \frac{138-28 \lambda +26 \lambda ^2-8 \sqrt{144+\lambda  (-112+\lambda  (113+9 \lambda  (-14+9 \lambda )))}}{7 (27+\lambda  (-14+23 \lambda ))}  \,,  \\
m^2 &=& \frac{1}{5 \lambda  (-11+13 \lambda  (-6+17 \lambda ))}  \lbrack 26-7 \sqrt{144+\lambda  (-112+\lambda  (113+9 \lambda  (-14+9 \lambda )))} \nonumber \\
&+& \lambda  \lbrace-1661+104 \sqrt{144+\lambda  (-112+\lambda  (113+9 \lambda  (-14+9 \lambda )))} \nonumber \\
 &+& \lambda  \left(2065+13 \lambda  (-79+9 \lambda )-13 \sqrt{144+\lambda  (-112+\lambda  (113+9 \lambda  (-14+9 \lambda )))}\right)\rbrace\rbrack \,,  \nonumber \\
\Lambda &=& \frac{1}{10 \lambda  (-11+13 \lambda  (-6+17 \lambda ))} \lbrack -463-34 \sqrt{144+\lambda  (-112+\lambda  (113+9 \lambda  (-14+9 \lambda )))}\nonumber \\
&-& 6 \lambda \left(357+52 \sqrt{144+\lambda  (-112+\lambda  (113+9 \lambda  (-14+9 \lambda )))}\right) \nonumber \\
&+ & \lambda^2  \left(10000+39 \lambda  (-166+81 \lambda )+754 \sqrt{144+\lambda  (-112+\lambda  (113+9 \lambda  (-14+9 \lambda )))}\right) \rbrack \,,  \nonumber \\
A_2 &=& \frac{1}{4} \left(8+\lambda  (-7+9 \lambda )+\sqrt{144+\lambda  (-112+\lambda  (113+9 \lambda  (-14+9 \lambda )))}\right) \,. \nonumber 
 \end{eqnarray}
 }
An acceptable solution with vanishing horizon area arises if  $\lambda>0$ and $\lambda \ne 1$, $\lambda \ne \frac{1}{221} \left(39+4 \sqrt{247}\right)$.
 \item {\bf Solution for $\alpha=-3$ and $\delta=-1$:}
 {\scriptsize
 \begin{eqnarray}
C_a^2 &=& \frac{5}{(25-39 \lambda )^2 (-1+\lambda ) (11+39 \lambda )^2} \lbrack
-12125 \left(-30+\sqrt{900+\lambda  (380+\lambda  (-1219+39 \lambda  (-38+39 \lambda )))}\right) \nonumber \\
&+& \lambda  \left(220485-4978 \sqrt{900+\lambda  (380+\lambda  (-1219+39 \lambda  (-38+39 \lambda )))} \right) \nonumber \\
&+& 39 \lambda^2  \left(-20395+5 \lambda  (-2075+3081 \lambda )+477 \sqrt{900+\lambda  (380+\lambda  (-1219+39 \lambda  (-38+39 \lambda )))}\right) \rbrack \,, \nonumber \\
\beta_{yz} &=& \frac{-380+3 \lambda  (-83+273 \lambda )+9 \sqrt{900+\lambda  (380+\lambda  (-1219+39 \lambda  (-38+39 \lambda )))}}{(-25+39 \lambda ) (11+39 \lambda )} \,, \nonumber \\
k &=& \frac{2 \left(185+\lambda  (-82+117 \lambda )-8 \sqrt{900+\lambda  (380+\lambda  (-1219+39 \lambda  (-38+39 \lambda )))}\right)}{(-25+39 \lambda ) (11+39 \lambda )} \,,   \\
m^2 &=& \frac{1}{5 \lambda  (-3395+109 \lambda  (-6+41 \lambda ))} \lbrack -2425 \left(37+\sqrt{900+\lambda  (380+\lambda  (-1219+39 \lambda  (-38+39 \lambda )))}\right) \nonumber \\
&+& \lambda  \left(44235-654 \sqrt{900+\lambda  (380+\lambda  (-1219+39 \lambda  (-38+39 \lambda )))}\right) \nonumber \\
&+& \lambda^2  \left(130567-327 \lambda  (217+34 \lambda )+3379 \sqrt{900+\lambda  (380+\lambda  (-1219+39 \lambda  (-38+39 \lambda )))}\right) \rbrack \,, \nonumber \\
\Lambda &=& \frac{1}
{10 \lambda (109 \lambda (41 \lambda-6)-3395)}  \lbrack {\lambda (\lambda (130567-327 \lambda (34 \lambda+217))+44235)-89725} \nonumber \\
&+& {(109 \lambda (31 \lambda-6)-2425) \sqrt{  
900+\lambda  (380+\lambda  (-1219+39 \lambda  (-38+39 \lambda )))
}} \rbrack \,, \nonumber \\
A_2 &=& \frac{1}{40} \left(-10+\lambda  (-19+39 \lambda )+\sqrt{900+\lambda  (380+\lambda  (-1219+39 \lambda  (-38+39 \lambda )))}\right)  \,. \nonumber 
 \end{eqnarray}
}
This solution is physically acceptable for all $\lambda>0$
except $\lambda =\frac{25}{39}, \frac{327+4 \sqrt{954949}}{4469},1$.
\end{itemize}

These two classes of solutions are quite messy, but are presented  (in a somewhat Pyrrhic victory) 
to illustrate that physically sensible solutions of this sort do exist for several values of $\alpha$, $\delta$.

\section{Striped phases by Kaluza-Klein reduction of 5d type VII$_0$}{\label{sec:stripe}}

By using the dimensional reduction we reviewed in \S {\ref{sec:KKred}},  
we can also obtain simple analytical examples of striped phases. 
For that purpose, we start with 5d type VII$_0$ case. 
See for examples of the other constructions of ``striped" phases in the literature \cite{holstripes1,
Donosstripes, holstripes2, holstripes3, Rozali,DGnew}.

Let's consider the 5d type VII$_0$ solution \cite{Bianchi}, which takes the following ansatz 
\begin{eqnarray}
d\hat s^2 &=& dr^2-e^{2\beta_t r} dt^2+ (dx^1)^2+ e^{2 \beta r} \left[ (c \, dx^2+ s \, dx^3)^2+\lambda^2 (-s \, dx^2+ c \, dx^3)^2\right] \,, \\
\hat A &=& \sqrt{\tilde A_2} e^{\beta r} (c \, dx^2+ s \, dx^3) \,,
\end{eqnarray}
where  $c=\cos x^1$ and $s=\sin x^1$. 
This is the solution of the 5d action given by eqn. (\ref{EHaction_V_L}), where
\begin{eqnarray}
 m^2 &=& \f{2(11+2 \lambda^2-10\lambda^4+3\lambda^6)}{\lambda^2 (5 \lambda^2-11)} \,, \\
\Lambda &=& \f{1}{50} \left(95
\lambda^2+\f{25}{\lambda^2}-\f{50}{\lambda^2-2}+\f{144}{5\lambda^2-11}-146\right) \,, \\
\beta &=& \sqrt{2} \sqrt{\f{2-3\lambda^2+\lambda^4}{5\lambda^2-11}} \,, \label{valparabeta}\\
\beta_t &=& \f{\lambda^2-3}{\sqrt{2}~ (\lambda^2-2)}
\sqrt{\f{2-3\lambda^2+\lambda^4}{5\lambda^2-11}} \,, \label{valparabetat}\\
\ti A_2&=& \f{2}{2-\lambda^2}-2 \,,
\end{eqnarray}
and $1\le \lambda <\sqrt{2}$.

In order to dimensionally reduce this 5d solution to 4d along $x^2$, 
we re-write the metric and gauge field as 
\begin{eqnarray}
 d\hat s^2 &=& dr^2-e^{2\beta_t r} dt^2+ (dx^1)^2 + e^{2 \beta r} f(x^1) (dx^3)^2+ e^{2 \beta r}  (c^2+\lambda^2 s^2) (dx^2+B_3 dx^3)^2 \,, \\
\hat A &=&  \left(\sqrt{\tilde A_2} e^{\beta r} s-\chi B_3  \right) dx^3 +\chi (dx^2+B_3 dx^3) \,,
\end{eqnarray}
where 
\bea
\label{B3chif}
B_3=\frac{(1-\lambda^2)cs}{c^2+\lambda^2 s^2} \quad \,, \quad   
\chi=\sqrt{\tilde A_2} e^{\beta r} c \quad \,, \quad   
f(x^1)=\frac{\lambda^2  
}{c^2+\lambda^2 s^2} \,.
\eea

This metric can be written in the form 
\begin{equation}
 d \hat s^2 = e^{2 \alpha_1 \phi(x)} ds^2+ e^{2 \alpha_2 \phi(x)} \left(dz+B_{\mu}(x) dx^{\mu}\right)^2
\end{equation}
where $\alpha_2=-2 \alpha_1 =-\frac{1}{\sqrt{3}}$ using eq.(\ref{alpha_2alpha_1byd}) and 
setting $d=4$.
The lower dimensional scalar field depends non-trivially on $x^1$ through 
$c = \cos x^1$ and $s = \sin x^1$, so it is clear that this can give geometries dual to striped phases.

The 4d action (after appropriate Weyl rescaling etc) is given by eq.(\ref{dimredaction}),
\begin{eqnarray}
 S &=&\int d^{4} x \sqrt{| g|}~
\lbrack  R -\frac{1}{2} (\partial \phi)^2-\frac{1}{4} e^{-6 \alpha_1 \phi} H^2
-\frac{1}{4} e^{-2 \alpha_1 \phi} F^{2}-\frac{1}{2} e^{4 \alpha_1 \phi} (\partial \chi)^2 \nonumber \\
&& -\frac{1}{4} m^2 (A-\chi B)^2 -\frac{1}{4} m^2 e^{6 \alpha_1 \phi} \chi^2 + \Lambda e^{2 \alpha_1 \phi}\rbrack~.
\end{eqnarray}
The 4d solution is  given by,
\begin{eqnarray}
ds^2 &=& e^{\beta r } g(x^1) \left( dr^2-e^{2\beta_t r} dt^2+ (dx^1)^2 + e^{2 \beta r}  f(x^1) (dx^3)^2 \right) \,, \\
g(x^1) &=& \sqrt{c^2+\lambda^2 s^2} \quad \,, \quad 
\phi = -\sqrt{3}\left( \beta r + \log g(x^1)  \right) \,, \\
A &=& \sqrt{\tilde A_2} e^{\beta r}  s   dx^3 \quad \,, \quad 
B =  B_3 dx^3~
\end{eqnarray}
where $B_3$, $\chi$, $f(x^1)$ are given in (\ref{B3chif}). 

If we perform the coordinate change $r=\frac{1}{ \beta_t+\beta} \log (\tilde r)$, 
the field configuration simplifies slightly to:
\begin{eqnarray}
ds^2 &=& g(x^1) \left(\frac{1}{(\beta_t+\beta)^2} \frac{d\tilde r^2}{\tilde r^{2 \gamma}}-\tilde r^{2 \gamma} dt^2+ \tilde r^{2 \delta} (dx^1)^2 + f(x^1) \tilde r^{6 \delta}(dx^3)^2 \right)  \,, \\
\phi &=& -2 \sqrt{3}\,  \delta \, \log \tilde r -\sqrt{3}\log g(x^1) \quad \,, \quad 
\chi = \sqrt{\tilde A_2} \tilde r^{2 \delta} c \,, \\
A &=& \sqrt{\tilde A_2} \tilde r^{2 \delta}  s  dx^3  \quad \,, \quad  
B =  B_3 dx^3 \,,  
\end{eqnarray}
where 
\bea 
2 \gamma=\frac{2 \beta_t +\beta}{\beta_t+\beta}  \quad \,, \quad 
2 \delta=\frac{\beta}{\beta_t+\beta} \,.
\eea  
The ``striped" structure of the solution is evident and is seen in both the scalar field 
and in the metric.

\section{Classification of four dimensional homogeneous spaces}~\label{sec:Bianchi4algebras}

We have seen so far that by various simple modifications of the Bianchi horizons, it is possible to 
exhibit analytic striped phases and hyperscaling violation in anisotropic phases.  The more ambitious
goal of \cite{Bianchi} was to classify homogeneous, anisotropic extremal horizons as a tractable starting point
for a more general classification.

Here, we try to extend the classification proposed in \cite{Bianchi}.  
The holographic dual of a four-dimensional
quantum field theory has four-dimensional spatial slices (including the ``radial" direction), 
and therefore a 
classification based on four-dimensional real Lie algebras seems more natural 
for static metrics.  This allows
for the possibility that the radial direction is more non-trivially intertwined 
with the ``field theory spatial" dimensions. 

The classification of \cite{Patera} yields 12 different classes of four-dimensional real Lie algebras (including
some indexed by continuous parameters), with
a variety of inequivalent embeddings of fixed three sub-algebras (corresponding to the field theory space)
into each.  Here, we give the data
corresponding to the four-dimensional algebras in \S \ref{sec:Bianchi4algebrasSub1}, and describe the classification of subalgebras in \S \ref{sec:Bianchi4algebrasSub2}.

\subsection{Real four-dimensional Lie algebras}~\label{sec:Bianchi4algebrasSub1}

A four-dimensional homogeneous space ${\cal H}$ has four linearly independent Killing vectors $e_i$ with
$i=1, \cdots 4$, which generate the isometries of ${\cal H}$.  These satisfy an algebra
\begin{equation}
[ e_i,e_j ] = C^{k}_{ij} e_k
\end{equation}
with $C^{k}_{ij}$ the structure constants of the related four-dimensional real Lie algebra.  
It also has four invariant one-forms
$\omega^i$.  The Lie derivatives of the $\omega^i$ along all the $e_j$ directions vanish, and the
$\omega^i$'s can be normalized to satisfy the relation
\begin{equation}
d\omega^i = {1\over 2} C^{i}_{jk} \omega^j \wedge \omega^k~.
\end{equation}
Here, we will list the structure constants of the 12 inequivalent four-dimensional algebras, as well as convenient choices
for the Killing vectors and one-forms.  The one-forms are particularly useful because a metric
written in terms of them
\begin{equation}
ds^2 = \cdots + \eta_{ij} \omega^i \otimes \omega^j 
\end{equation}
(with $\cdots$ independent of the relevant four dimensions) will be invariant under the isometry group.

In the following, we adopt the notation of \cite{Patera} in naming the algebras -- we call them
$A_{4,k}$ with $k=1,\cdots,12$.  We list only the
non-vanishing structure constants (up to obvious permutation of indices).
\\
\begin{itemize}
 \item {\bf $A_{4,1}$}: $C^{1}_{24} = 1,~C^{2}_{34} = 1$\\
\begin{tabular}{cccc}
$e_1 = \partial_1$ & $e_2 = \partial_2$ & $e_3 = \partial_3$ & $e_4 = x_2 \partial_1 + x_3 \partial_2 + \partial_4$\\
$\omega^1 = dx^1 - x_4 dx^2 + {1\over 2} x_4^2 dx^3 $ & $\omega^2 = dx^2 - x_4 dx^3$ & $\omega^3 = dx^3$
& $\omega^4 = dx^4$
\end{tabular}
\\
\\

\item {\bf $A^a_{4,2}$}: $C^1_{14}= a,~C^2_{24}=1, ~C^{2}_{34}=1,~C^{3}_{34} = 1$ \\
\begin{tabular}{cccc}
$e_1 = \partial_1$ & $e_2 = \partial_2$ & $e_3 = \partial_3$ & $e_4 = a x_1 \partial_1 + (x_2 + x_3) \partial_2 + x_3 \partial_3+\partial_4$\\
$\omega^1 = e^{-ax_4} dx^1$ & $\omega^2 = e^{-x_4} (dx^2 - x_4 dx^3)$ & $\omega^3 = e^{-x_4} dx^3$ &
$\omega^4 = dx^4$
\end{tabular}
\\
\\

\item {\bf $A_{4,3}$}: $C^1_{14}=1,~C^{2}_{34}=1$ \\
\begin{tabular}{cccc}
$e_1 = \partial_1$ & $e_2 = \partial_2$ & $e_3 = \partial_3$ & $e_4 = x_1 \partial_1 + x_3 \partial_2 + \partial_4$\\
$\omega^1 = e^{-x_4} dx^1$ & $\omega^2 = dx^2 - x_4 dx^3$ & $\omega^3 = dx^3$ & $\omega^4 = dx^4$
\end{tabular}
\\
\\

\item {\bf $A_{4,4}$}: $C^{1}_{14}=1,~C^{1}_{24}=1,~C^{2}_{24}=1,~C^{2}_{34}=1,~C^{3}_{34}=1$\\
\begin{tabular}{cccc}
$e_1 = \partial_1$ & $e_2 = \partial_2$ & $e_3 = \partial_3$ & \\ $e_4 = (x_1 + x_2)\partial_1 + (x_2 + x_3)\partial_2 + x_3 \partial_3 + \partial_4$\\& & &\\
$\omega^1 = e^{-x_4} (dx^1 - x_4 dx^2 + {1\over 2}x_4^2 dx^3)$ & $\omega^2 = e^{-x_4} (dx^2 - x_4 dx^3)$
& $\omega^3 = e^{-x_4} dx^3$ &$ \omega^4 = dx^4$
\end{tabular}
\\
\\

\item {\bf $A_{4,5}^{a,b}$}: $C^1_{14}=1,~C^2_{24}=a,~C^{3}_{34} = b$\\
\begin{tabular}{cccc}
$e_1 = \partial_1$ & $e_2 = \partial_2$ & $e_3 = \partial_3$ & $e_4 = x_1 \partial_1 + a x_2 \partial_2 + b x_3 \partial_3 + \partial_4$\\
$\omega^1 = e^{-x_4} dx^1$ & $\omega^2 = e^{-ax_4} dx^2$ & $\omega^3 = e^{-bx_4} dx^3$ &
$\omega^4 = dx^4$
\end{tabular}
\\
\\

\item {\bf $A_{4,6}^{a,b}$}: $C^1_{14}=a,~C^{2}_{24}=b,~C^3_{24}=-1,~C^2_{34} = 1,~C^3_{34}=b$\\
\begin{tabular}{cccc}
$e_1 = \partial_1$ & $e_2 = \partial_2$ & $e_3 = \partial_3$ &\\
 $e_4 = ax_1 \partial_1 + (bx_2 + x_3) \partial_2 + (bx_3 - x_2) \partial_3 + \partial_4$&&&\\
$\omega^1 = e^{-ax_4} dx^1$ & $\omega^2 = e^{-bx_4}[cos(x_4) dx^2 - sin(x_4) dx^3]$ & &\\
$\omega^3 = e^{-bx_4} (cos(x_4)dx^3 + sin(x_4) dx^2)$&
  $\omega^4 = dx^4$&&
\end{tabular}
\\
\\

\item {\bf $A_{4,7}$}: $C^1_{14}=2,~C^2_{24}=1,~C^{2}_{34}=1,~C^{3}_{34}=1,~C^{1}_{23}=1$\\
\begin{tabular}{cccc}
$e_1 = \partial_1$ & $e_2 = \partial_2 - {1\over 2}x_3 \partial_1$ & $e_3 = \partial_3 + {1\over 2} x_2 \partial_1$
& \\
$e_4 = 2x_1 \partial_1 + (x_2 + x_3) \partial_2 + x_3 \partial_3 + \partial_4$ &&&\\
$\omega^1 = e^{-2x_4}( dx^1 + {1\over 2} x_2 dx^3 - {1\over 2} x_3 dx^2)$&$\omega^2 = e^{-x_4} (dx^2 - x_4 dx^3)$&&\\
$\omega^3 = e^{-x_4}dx^3$&$\omega^4 = dx^4$&&
\end{tabular}
\\
\\

\item {\bf $A_{4,8}$}: $C^1_{23}=1,~C^2_{24}=1,~C^{3}_{34}=-1$\\
\begin{tabular}{cccc}
$e_1 = \partial_1$ & $e_2 = \partial_2 - {1\over 2}x_3 \partial_1$ & $e_3 = \partial_3 + {1\over 2} x_2 \partial_1$
& \\
$e_4 =x_2 \partial_2 - x_3 \partial_3 + \partial_4$ &&&\\
$\omega^1 = dx^1 + {1\over 2} x_2 dx^3 - {1\over 2} x_3 dx^2$ & $\omega^2 = e^{-x_4} dx^2$ &
$\omega^3 = e^{x_4} dx^3$ & $\omega^4 = dx^4$
\end{tabular}
\\
\\

\item{\bf $A_{4,9}^{b}$}: $C^{1}_{23}=1,~C^{1}_{14} = 1+b,~C^{2}_{24}=1,~C^3_{34}=b$\\
\begin{tabular}{cccc}
$e_1 = \partial_1$ & $e_2 = \partial_2 - {1\over 2} x_3 \partial_1$ & $e_3 = \partial_3 + {1\over 2} x_2 \partial_1$&\\
 $e_4 = (1+b) x_1 \partial_1 + x_2 \partial_2 + bx_3 \partial_3 + \partial_4$&&&\\
$\omega^1 = e^{-(b+1)x_4}(dx^1 + {1\over 2} x_2 dx^3 - {1\over 2} x_3 dx^2)$ & $\omega^2 = e^{-x_4} dx^2$
&$ \omega^3 = e^{-bx_4} dx^3$ & $\omega^4 = dx^4$
\end{tabular}
\\
\\

\item{\bf $A_{4,10}$}: $C^{1}_{23}=1,~C^{3}_{24} = -1,~C^{2}_{34}=1$\\
\begin{tabular}{cccc}
$e_1 = \partial_1$ & $e_2 = \partial_2 - {1\over 2} x_3 \partial_1$ & $e_3 = \partial_3 + {1\over 2} x_2 \partial_1$&\\$e_4 = -x_2 \partial_3 + x_3 \partial_2 + \partial_4$\\
$\omega^1 = dx^1 + {1\over 2} x_2 dx^3 - {1\over 2} x_3 dx^2$&
$\omega^2 = cos(x_4) dx^2 - sin(x_4) dx^3$&$\omega^3 = cos(x_4) dx^3 + sin(x_4) dx^2$&\\
$\omega^4 = dx^4$&&&
\end{tabular}
\\
\\

\item{\bf $A_{4,11}^a$}: $C^{1}_{23}=1,~C^{1}_{14} = 2a,~C^{2}_{24}=a,~C^{3}_{24}=-1,~C^{2}_{34}=1,~C^{3}_{34}=a$\\
\begin{tabular}{cccc}
$e_1 = \partial_1$ & $e_2 = \partial_2 - {1\over 2} x_3 \partial_1$ &&\\
 $e_3 = \partial_3 + {1\over 2} x_2 \partial_1$& $e_4 = 2 a x_1 \partial_1 + (ax_2 + x_3) \partial_2 + (ax_3 - x_2) \partial_3 + \partial_4$\\
$\omega^1 = e^{-2ax_4}(dx^1 + {1\over 2} x_2 dx^3 - {1\over 2} x_3 dx^2)$&
$\omega^2 = e^{-ax_4}(cos(x_4) dx^2 - sin(x_4) dx^3)$&&\\
$\omega^3 = e^{-ax_4}(cos(x_4) dx^3 + sin(x_4) dx^2)$&$\omega^4 = dx^4$&&
\end{tabular}
\\
\\

\item{\bf $A_{4,12}$}: $C^{1}_{13}=1,~C^{2}_{23} = 1,~C^{2}_{14}=-1,~C^{1}_{24}=1$\\
\begin{tabular}{cccc}
$e_1 = \partial_1$&$e_2 = \partial_2$ & &\\
$e_3 = \partial_3 + x_1 \partial_1 + x_2 \partial_2$&$e_4 = \partial_4 + x_2 \partial_1 - x_1 \partial_2$&&\\
$\omega^1 = e^{-x_3}(cos(x_4) dx^1 - sin(x_4) dx^2)$&
$\omega^2 = e^{-x_3}(cos(x_4) dx^2 + sin(x_4) dx^1)$&&\\
$\omega^3 = dx^3$&$\omega^4 = dx^4$&&
\end{tabular}
\end{itemize}

\subsection{Three-dimensional subalgebras}~\label{sec:Bianchi4algebrasSub2}

We would like the bulk geometry to reflect homogeneity of the spatial slices in the dual field theory.
For this to happen, we wish to embed a three-dimensional real Lie algebra $A_3 \subset A_{4,k}$. The associated Killing vectors will generate isometries
of the spatial dimensions in the dual field theory; the non-trivial embedding of $A_3$ in $A_4$ 
reflects the intertwining of the ``spatial" and ``radial" directions along the flow, which allows us to
generalize the solutions of 
\cite{Bianchi} , where the full four-dimensional algebra was semi-simple and contained a trivial factor
corresponding to scale transformations.

These subalgebras (including inequivalent embeddings of a fixed $A_3$ into a given $A_{4,k}$) have
been classified in \cite{Patera}.  The results are as follows.  We name the subalgebras following the
convention of \cite{Patera}\footnote{For a dictionary relating the notation of \cite{Patera} for three-algebras with the more standard Bianchi nomenclature \cite{Landau}, see
Appendix \ref{Appendix:PateraBianchiDict}.}.  We also list the generators after each subalgebra.
\\
\\
\begin{itemize}
 \item {\bf $A_{4,1}$}: 
\begin{tabular}{cc}
$3A_1$&$e_1, e_2, e_4$\\
$A_{3,1}$&$e_4 + xe_3, e_2, e_1$\\
\end{tabular}
\\
\\
\item {\bf $A_{4,2}^a,$ $a \neq 0,1$}:
\begin{tabular}{cc}
$3A_1$&$e_1, e_2, e_3$\\
$A_{3,2}$&$e_4, e_2, e_3$\\
$A_{3,4}$&$e_4 ,e_1, e_2$ ~$(a=-1)$\\
$A_{3,5}^{z}$&$e_4, e_1, e_2$, $z=a (a<1)$, $z=1/a (a>1)$\\
\end{tabular}
\\
\\
\item {\bf $A_{4,2}^1$}:
\begin{tabular}{cc}
$3A_1$&$e_1, e_2, e_3$\\
$A_{3,2}$&$e_4, e_2, e_3 + x e_1$\\
$A_{3,3}$&$e_4, e_1, e_2$\\
\end{tabular}
\\
\\
\item {\bf $A_{4,3}$}:
\begin{tabular}{cc}
$3A_1$&$e_1, e_2, e_3$\\
$A_2 \oplus A_1$&$e_4 + xe_3, e_1; e_2$\\
$A_{3,1}$&$e_3, e_4, e_2$\\
\end{tabular}
\\
\\
\item {\bf $A_{4,4}$}:
\begin{tabular}{cc}
$3A_1$&$e_1, e_2, e_3$\\
$A_{3,2}$&$e_4, e_1, e_2$\\
\end{tabular}
\\
\\
\item {\bf $A_{4,5}^{a,b}$ $(-1 \leq a < b < 1, ab \neq 0)$}:
\begin{tabular}{cc}
$3A_1$&$e_1, e_2, e_3$\\
$A_{3,5}^a$&$e_4, e_1, e_2$\\
$A_{3,5}^b$&$e_4, e_1, e_3$\\
$A_{3,5}^z$&$e_4, e_2, e_3$ ~$z = a/b, |a/b| < 1; z = b/a, |a/b|>1$\\
\end{tabular}
\\
\\
\item {\bf $A_{4,5}^{a,a}$ $(-1 \leq a < 1, a \neq 0)$}:
\begin{tabular}{cc}
$3A_1$&$e_1, e_2, e_3$\\
$A_{3,3}$&$e_4, e_2, e_3$\\
$A_{3,5}^a$&$e_4, e_1, e_2 cos(\phi) + e_3 sin(\phi)$\\
\end{tabular}
\\
\\
\item {\bf $A_{4,5}^{a,1}$ $(-1 \leq a < 1, a \neq 0)$}:
\begin{tabular}{cc}
$3A_1$&$e_1, e_2, e_3$\\
$A_{3,3}$&$e_4, e_1, e_3$\\
$A_{3,5}^a$&$e_4, e_1 cos(\phi) + e_3 sin(\phi), e_2$\\
\end{tabular}
\\
\\
\item {\bf $A_{4,5}^{1,1}$}:
\begin{tabular}{cc}
$3A_1$&$e_1 e_2, e_3$\\
$A_{3,3}$&$e_4, e_1 + xe_3, e_2 + ye_3$\\
$$&$e_4, e_1 + xe_2, e_3$\\
$$&$e_4, e_2, e_3$\\
\end{tabular}
\\
\\
\item {\bf $A_{4,6}^{a,b}$ $(a \neq 0, b>0)$}:
\begin{tabular}{cc}
$3A_1$&$e_1, e_2, e_3$\\
$A_{3,7}^b$&$e_4, e_2 ,e_3$\\
\end{tabular}
\\
\\
\item {\bf $A_{4,7}$}:
\begin{tabular}{cc}
$A_{3,1}$&$e_2, e_3, e_1$\\
$A_{3,5}^{1\over 2}$&$e_4, e_1, e_2$\\
\end{tabular}
\\
\\
\item {\bf $A_{4,8}$}:
\begin{tabular}{cc}
$A_{3,1}$&$e_2, e_3, e_1$\\
$A_2 \oplus A_1$&$e_4, e_2; e_1$\\
$ $& $e_4, e_3; e_1$\\
\end{tabular}
\\
\\
\item {\bf $A_{4,9}^{b}$ $(0 < |b| < 1)$}:
\begin{tabular}{cc}
$A_{3,1}$&$e_2, e_3, e_1$\\
$A_{3,5}^z$&$e_4, e_1, e_2$ $z= 1 + b, |1+b| < 1; z = {1\over {1+b}}, |1+b| > 1$\\
$A_{3,4}$&$e_4, e_1, e_3$ $b=-{1\over 2}$\\
$A_{3,5}^z$&$e_4, e_1, e_3$ $z = {b\over {1+b}}, |{b \over {1+b}}| < 1; z = {{1+b} \over b}, |{{1+b}\over b}| > 1$\\
\end{tabular}
\\
\\
\item {\bf $A_{4,9}^1$}:
\begin{tabular}{cc}
$A_{3,1}$&$e_2, e_3, e_1$\\
$A_{3,5}^{1\over 2}$&$e_4, e_1, e_2 cos(\phi) + e_3 sin(\phi)$\\
\end{tabular}
\\
\\
\item {\bf $A_{4,9}^0$}:
\begin{tabular}{cc}
$A_{3,1}$&$e_2, e_3, e_1$\\
$A_2 \oplus A_1$&$e_1, e_4; e_3$\\
$A_{3,3}$&$e_4, e_1, e_2$\\
$A_{3,2}$&$e_4 + xe_3, e_1, e_2$ $x \neq 0$
\end{tabular}
\\
\\
\item {\bf $A_{4,10}$}:
\begin{tabular}{cc}
$A_{3,1}$&$e_2, e_3, e_1$\\
\end{tabular}
\\
\\
\item {\bf $A_{4,11}$ $(0 < a)$}:
\begin{tabular}{cc}
$A_{3,1}$&$e_2, e_3, e_1$\\
\end{tabular}
\\
\\
\item {\bf $A_{4,12}$}:
\begin{tabular}{cc}
$A_{3,3}$&$e_3, e_1, e_2$\\
$A_{3,6}$&$e_4, e_1, e_2$\\
$A_{3,7}^{|x|}$&$e_4 + xe_3, e_1, e_2$ $x \neq 0$
\end{tabular}

\end{itemize}

\section{Null energy condition for 5d space-times 
with four-algebras}{\label{sec:NECfor5d}}

We have proposed a classification of extremal near-horizon metrics for 5d black branes using
four-algebras.
The first check of validity
for space-times which realize these symmetries is whether they can be supported by reasonable
matter content, i.e. whether the stress energy tensor that supports
the space-time satisfies the Null Energy Condition (NEC):
\begin{equation}
T_{\mu\nu}N^{\mu}N^{\nu}\geq0
\end{equation}
for all future directed null vectors $N^{\mu}.$ Via Einstein's equations
this translates into a constraint on the geometry, namely that the Einstein
tensor has to satisfy

\begin{equation}
\label{nullEconditionforarbitraryNN}
G_{\mu\nu}N^{\mu}N^{\nu}\geq 0 \,.
\end{equation}

We will work in an abstract orthonormal basis, in which the metric takes the
form 
\bea
ds^{2}=\sum_{i=1}^{4} (\sigma^{i})^2 -(\sigma^{t})^2 \,,
\eea 
for $\sigma^{i} \equiv \lambda_{ij}\omega^{j}$,
where $\omega^{j}$ are the invariant one-forms listed before. For
simplicity we shall only take the matrix $\lambda_{ij}$ to be diagonal:
$\lambda_{ij}=\lambda_{i}\delta_{ij}$ and $\sigma^i = \lambda_i \omega^i $ (no sum for $i$). 
The time-like one-form is 
given by $\sigma^{t}=\sqrt{g_{tt}(r)}dt$, where we will only consider
a simple scaling red-shift factor: $g_{tt}(r)=e^{2\beta_{t}r}$. The
radial coordinate $r$ is identified with one of the exact one-forms
$\lambda_i dr=\sigma^{i}$ such that the other 3 one-forms form a sub-algebra. 
In this basis an arbitrary future-directed null vector can easily
be written by using 4 real parameters $s_i$ ($i=1,2,3,4$) in the form:

\begin{equation}\label{generalnullvector}
\vec{N}=(\sum_{i=1}^{4} s_{i}^{2})^{\frac{1}{2}} X_{t}+\sum_{i=1}^{4} s_{i} X_{i}~
\end{equation}
where $X_t$, $X_i$ are dual vectors to the invariant one-forms $\sigma^t$, $\sigma^i$. 
Equation (\ref{nullEconditionforarbitraryNN}) then becomes a bilinear function 
\bea
\label{definitionofmatrixM}
G_{\mu\nu}N^{\mu}N^{\nu} \equiv M_{ij} s^{i} s^{j} \ge 0 \,,
\eea
for arbitrary choices of the four real parameters $s^{i}, i=1,2,3,4$. Since the NEC amounts to imposing
positive definiteness of this bilinear function, it then requires
that the eigenvalues of the matrix $M_{ij}$ must be non-negative.

We will study natural multiparameter families of metrics realizing a given symmetry structure.
We conclude that if a given set of parameters violates the NEC, it cannot be supported by physically
reasonable matter fields.  In this case, we view the metric with those parameters as capturing a physically unrealizable geometry.  In some cases the entire ``natural" set of possibilities for realizing
a given four-algebra can be eliminated -- in this case, we conclude that the related type is not realized
physically.

In the following we will study the NEC for each of
the 12 types of space-times (in a natural metric parametrization realizing the associated four-algebra), by writing out specifically the eigenvalues
of the $M_{ij}$, and analyzing the constraints obtained by requiring that
all of these eigenvalues be non-negative. 

Before listing the results of applying the NEC to each type, let's discuss the constraint on $\beta_{t}, \beta_i$.
In general, a structure coefficient of the form $C_{ir}^{i}=k_{i}$
signifies a scaling of the form $e^{-k_{i}r}$ for the one-form $\sigma^{i}$.
The red-shift factor $g_{tt}(r)$ is $e^{2\beta_{t}r}$, and the limit where the red-shift
factor $g_{tt}$ approaches zero corresponds to the IR in the dual theory.  
A field theory spatial volume 
$\sim e^{-\sum_{i=1}^{3} {k_{i}} r} $
in the IR limit corresponds to a ground state entropy density of the field theory.
In order to satisfy the laws of thermodynamics, we require that the field theory 
entropy density goes to zero in the IR. 
%This condition was also stated in the earlier footnote 3. %\footnoteref{PhysicallyAcceptable}.
On the gravity side, this means that the spatial volume $\sim e^{-\sum_{i=1}^{3} {k_{i}} r} $ 
should have the same qualitative scaling behavior as the red-shift factor, in the sense that
\bea
{\rm {\bf Nernst's~Law}}: \beta_{t}\sum_{i=1}^{3}C_{ir}^{i}<0 \,.
\eea
We will impose this, as well as the NEC, as a criterion of reasonableness, which we'll call
``Nernst's law."\footnote{As already mentioned in some of the discussion above  this may in fact 
be too strong.  For instance, $AdS_2 \times R^d$ near-horizon geometries, which violate this criterion, 
are ubiquitous in simple approximations to string theory, and show fascinating physical
properties.  However, these have $k_i = 0$, and one could contemplate softening the 
condition above to $\leq$, at least in some cases.}
 
We will now study the NEC separately for each of the $A_{4,k}$ cases, where the natural metric ansatz we
will consider is 
\bea
ds^2 = - e^{2\beta_{t}r} dt^2 + \sum_{i=1}^4 \lambda_i^2 (\omega^i)^2 \,.  
\eea

\begin{itemize}
\item $A_{4,1}:$ We identify $x^4$ as a radial coordinate;  
$\lambda_4 dr = \sigma^4 = \lambda_4 \omega^4$ gives $x^4 = r$.

It is straightforward to calculate the matrix $M$ defined by eq.(\ref{definitionofmatrixM}). 
One of the eigenvalues of the matrix $M$ becomes  
\bea
M_1 = -\frac{\lambda_1^2 \lambda_3^2+\lambda_2^4}{2 \lambda_2^2 \lambda_3^2
   \lambda_4^2} < 0 \,,
\eea
which is negative. Therefore this geometry is ruled out. 

\item $A_{4,2}^{a}:$ We identify $x^4$ as a radial coordinate ($x^4 = r$). 
The eigenvalues of $M$ are
\bea
M_1 &=& -\frac{(a+2 -\beta_t) (a+\beta_t)}{\lambda_4^2}  \,, \quad  
M_2 = -\frac{2 \lambda_3^2 \left(a^2+(a+2)
   \beta_t+2\right)+\lambda_2^2}{2 \lambda_3^2 \lambda_4^2} \,, \\
M_{3,4} &=& \frac{(\beta_t+1)
   (-a-2 +\beta_t)}{\lambda_4^2}
   \pm \frac{\sqrt{\lambda_2^2 \lambda_3^4 \lambda_4^4
   \left(\lambda_3^2 (a+2 -\beta_t)^2+\lambda_2^2\right)}}{2 \lambda_3^4
   \lambda_4^4} \,. 
\eea

Nernst's law forces 
\bea
\beta_t (a + 2) < 0 \,.
\eea
It can be shown that 
the NEC and Nernst's law are satisfied in several open sets of
parameter space; for example: 
with all $\lambda_i = O(1)$ for $i=1,2,3,4$, it is easy to see that at large negative values of $\beta_t$ with $-2 < a$, 
 all conditions are satisfied.   

\item $A_{4,3}:$ We identify $x^4$ as a radial coordinate ($x^4 = r$).   
The eigenvalues of $M$ are
\bea
M_1&=& \frac{\beta_t^2-1}{\lambda_4^2} \,, \quad 
M_2= -\frac{2 \beta_t+\frac{\lambda_2^2}{\lambda_3^2}+2}{2
   \lambda_4^2} \,, \\
M_{3,4}&=&   \frac{2 (\beta_t-1) \beta_t \lambda_3^4
   \lambda_4^2 \pm \sqrt{\lambda_2^2 \lambda_3^4 \lambda_4^4
   \left((\beta_t-1)^2 \lambda_3^2+\lambda_2^2\right)}}{2 \lambda_3^4 \lambda_4^4} \,. 
\eea
Nernst's law forces $\beta_t < 0$. 
The NEC can be easily satisfied at large negative values of $\beta_t$. 

\item $A_{4,4}:$ We identify $x^4$ as a radial coordinate ($x^4 = r$).   
The eigenvalues of $M$ are
\bea
M_1 = -\frac{\frac{\lambda_1^2}{\lambda_2^2}+\frac{\lambda_2^2}{\lambda_3^2}+6 + 6 \beta_t}{2
   \lambda_4^2}
\eea
The other three eigenvalues $M_{2,3,4}$ are roots of a higher order polynomial,
which is too cumbersome to write down. 
However in the large negative limit $\beta_t \to -\infty$, we can easily check  that the rest 
$M$ eigenvalues behave as 
\bea
M_{2,3,4} \to \frac{\beta_t^2}{\lambda_4^2 } + O(\beta_t) \,.  
\eea
On the other hand, Nernst's law forces $\beta_t < 0$. 
Therefore the NEC can be easily  satisfied in the regime with large negative values of $\beta_t$.

\item $A_{4,5}^{a,b}:$ We identify $x^4$ as a radial coordinate ($x^4 = r$).  
The eigenvalues of $M$ are
\bea
M_{1}&=& -\frac{(1+ \beta_t) (a+b-\beta_t+1)}{\lambda_4^2} \,, \quad 
M_2 = -\frac{(a+\beta_t)  (a+b-\beta_t+1)}{\lambda_4^2} \,, \\ 
M_3 &=&  -\frac{(b+\beta_t) (a+b-\beta_t+1)}{\lambda_4^2} \,, \quad 
M_4 = -\frac{1+a^2+b^2+ \beta_t (1 + a + b) }{\lambda_4^2} \,.
\eea
Nernst's law forces 
\bea
\beta_t (1 + a + b) < 0 \,.
\eea
We see that there is an open parameter set satisfying both the NEC and Nernst's law 
for large negative $\beta_t$ with $1 + a + b > 0$.

\item $A_{4,6}^{a,b}:$ We identify $x^4$ as a radial coordinate ($x^4 = r$).  
The eigenvalues of $M$ are
\bea
M_1 & = & -\frac{(a+\beta_t) (a+2 b-\beta_t)}{\lambda_4^2} \,, \\ 
M_2&=& -\frac{2 a^2+2 a
   \beta_t+4 b
   (b+\beta_t)+\frac{\lambda_3^2}{\lambda_2^2}+\frac{\lambda_2^2}{\lambda_3^2}-2}{2 \lambda_4^2} \,, \\ 
M_{3,4} &=& -\frac{(b+\beta_t) (a+2
   b-\beta_t)}{\lambda_4^2} \nonumber \\
 &&   \pm  
   \frac{\sqrt{\lambda_2^4 \lambda_4^4
   \left(\lambda_3^4-\lambda_2^2 \lambda_3^2\right)^2 \left(\lambda_2^2
   \lambda_3^2 \left((a+2
   b-\beta_t)^2+2\right)+\lambda_2^4+\lambda_3^4\right)}}{2 \lambda_2^4
   \lambda_3^4 \lambda_4^4}~.
\eea
Nernst's law forces 
\bea
\beta_t (a + 2 b) < 0 \,.
\eea
We see that there is an open parameter set satisfying both the NEC and Nernst's law 
for large negative $\beta_t$ with $a + 2 b > 0$.

\item $A_{4,7}:$ We identify $x^4$ as a radial coordinate ($x^4 = r$).   
The eigenvalues of $M$ are
\bea
M_1 &=&  \frac{\beta_t^2 - 2 \beta_t - 8}{\lambda_4^2}
   +\frac{\lambda_1^2}{2 \lambda_2^2\lambda_3^2}
   \,, \quad 
M_2 =   -\frac{8 \beta_t+\frac{\lambda_2^2}{\lambda_3^2}+12}{2
   \lambda_4^2}  \,, \\
M_{3,4}&=&  \frac{(\beta_t-4)
   (\beta_t+1)}{\lambda_4^2}-\frac{\lambda_1^2}{2 \lambda_2^2
   \lambda_3^2}
  \pm  \frac{\sqrt{\lambda_2^{10} \lambda_3^4 \lambda_4^4
   \left((\beta_t-4)^2 \lambda_3^2+\lambda_2^2\right)}}{2 \lambda_2^4
   \lambda_3^4 \lambda_4^4} \,.
\eea
Nernst's law gives 
\bea
\beta_t < 0 \,.
\eea
It is obvious that there is an open parameter range satisfying the NEC and Nernst's law at large negative $\beta_t$.

\item $A_{4,8}:$ We identify $x^4$ as a radial coordinate ($x^4 = r$).  
The eigenvalues of $M$ are
\bea
M_1&=& -\frac{2}{\lambda_4^2} \,, \quad M_2 = \frac{\beta_t^2}{\lambda_4^2}+\frac{\lambda_1^2}{2
   \lambda_2^2 \lambda_3^2} \,, \\
M_{3,4} &=&  -\frac{\lambda_1^2}{2  \lambda_2^2   \lambda_3^2} + \frac{ \beta_t (\beta_t \pm 1) }{\lambda_4^2} \,.
\eea
We find that the matrix $M$ has one negative eigenvalue, $M_{1}=-\frac{2}{\lambda_4^2}$,
hence this type of metric is ruled out by the NEC. 

\item $A_{4,9}^{b}:$ We identify $x^4$ as a radial coordinate ($x^4 = r$).  
The eigenvalues of $M$ are
\bea
M_1 &=& - \frac{ (2 b- \beta_t + 2) (1+ \beta_t) }{\lambda_4^2}
-\frac{\lambda_1^2}{2 \lambda_2^2  \lambda_3^2} \,, \\
M_2 &=& -\frac{(2 b-\beta_t+2)
   (b+\beta_t)}{\lambda_4^2}-\frac{\lambda_1^2}{2 \lambda_2^2
   \lambda_3^2}  \,,\\ 
M_3 &=&-\frac{(2 b-\beta_t+2) (b+\beta_t+1)}{\lambda_4^2} 
+ \frac{\lambda_1^2}{2 \lambda_2^2 \lambda_3^2} \,, \\
M_4&=& -\frac{2 (b   (b+\beta_t+1)+\beta_t+1)}{\lambda_4^2}  \,.
\eea
Nernst's law forces 
\bea
\beta_t (b + 1) < 0 \,.
\eea
Therefore, the NEC and Nernst's law are satisfied on an open set where
 $b > -1$ and large negative $\beta_t$. 

\item $A_{4,10}:$ We identify $x^4$ as a radial coordinate ($x^4 = r$).  
The eigenvalues of $M$ are
\bea
M_2 &=& 
 -\frac{\left(\lambda_2^2-\lambda_3^2\right)^2}{2 \lambda_2^2  \lambda_3^2 \lambda_4^2} 
\,, \quad 
M_1 =
\frac{\beta_t^2}{\lambda_4^2}+\frac{\lambda_1^2}{2 \lambda_2^2 \lambda_3^2}
\,, \\
 M_{3,4} &=&  \frac{2 \beta_t^2 \lambda_2^4 \lambda_3^4
   \lambda_4^2    -\lambda_1^2 \lambda_2^2  \lambda_3^2 \lambda_4^4 
   \pm \sqrt{\lambda_2^4 \lambda_4^4
   \left(\lambda_3^4-\lambda_2^2 \lambda_3^2\right)^2
   \left(\left(\beta_t^2+2\right) \lambda_2^2
   \lambda_3^2+\lambda_2^4+\lambda_3^4\right)}}{2 \lambda_2^4 \lambda_3^4
   \lambda_4^4} \,. \quad  \quad 
\eea
Notice that $M_{1}$ is negative and hence violates the NEC unless $\lambda_2=\lambda_3$. 
However, in the case where $\lambda_2=\lambda_3$, we have the metric 
\bea
ds^2 &=& -e^{2 \beta_t r} dt^2 + \lambda_4^2 dr^2 + (\lambda_1)^2  (\omega^1)^2 
+  \lambda_2^2 ( (\omega^2)^2 +  (\omega^3)^2)
\eea
where 
\bea
\label{thissubclasses}
\omega^1 = dx^1 + \frac12 x^2 dx^3 - \frac12 x^3 dx^2 \,, \quad \omega^2 = dx^2 \,, \quad \omega^3 = dx^3 \,.
\eea
This is a type II Bianchi geometry\footnote{One can easily check that basis 
(\ref{thissubclasses}) gives 
$d \omega^i = \frac{1}{2} C^i_{jk} \omega^j \wedge \omega^k$ with $C^1_{23} = - C^1_{32} = 1$ and the rest $C^i_{jk} = 0$.} of the sort studied in \S 4.1 of \cite{Bianchi}.
Therefore, the generic case of $A_{4,10}$ with $\lambda_2 \neq \lambda_3$ is ruled out by the NEC. 
In the case $\lambda_2 = \lambda_3$, the NEC imposes the constraint 
$\beta_{t}^{2} >  {\lambda_1^2 \lambda_4^2}/{2 \lambda_2^4}$.

\item $A_{4,11}^{a}:$ We identify $x^4$ as a radial coordinate ($x^4 = r$).  
The eigenvalues of $M$ are
\bea
M_1 &=& \frac{-8 a^2-2 a   \beta_t+\beta_t^2}{\lambda_4^2}
   +\frac{\lambda_1^2}{2 \lambda_2^2  \lambda_3^2} \,, \quad    
M_2 = -\frac{12 a^2+8 a
   \beta_t+\frac{\lambda_3^2}{\lambda_2^2}+\frac{\lambda_2^2}{
   \lambda_3^2}-2}{2 \lambda_4^2} \,, \quad \\
M_{3,4} &=&   - \frac{2 \lambda_2^2 \lambda_3^2  (4 a-\beta_t) (a+\beta_t)
      + \lambda_1^2 }{2 \lambda_2^2 \lambda_3^2  \lambda_4^2} \nonumber \\
&&    \pm \frac{\sqrt{\lambda_2^4
   \lambda_4^4 \left(\lambda_3^4-\lambda_2^2 \lambda_3^2\right)^2
   \left(\lambda_2^2 \lambda_3^2 \left((\beta_t-4
   a)^2+2\right)+\lambda_2^4+\lambda_3^4\right)}}{2 \lambda_2^4 \lambda_3^4
   \lambda_4^4} \,.
\eea
Nernst's law forces 
\bea
a \beta_t < 0 \,.
\eea
Therefore, there is an open parameter range for $a > 0$ and large negative $\beta_t$ where 
both the NEC and Nernst's law are satisfied. 

%%%%%%%%%%%%%%%%%%%%%%%%%%%%%%%%%%%%%%%%%%%%

\item $A_{4,12}:$ in this class both $x^{3}$ and $x^{4}$ are
obvious candidates for the radial coordinate.  

{\it Case 1: $r  = x^4$}

If we pick $r=x^{4}$, the eigenvalues of $M$ are 
\bea
M_1 &=&  -\frac{\left( \lambda_1^2 - \lambda_2^2\right)^2}{2 \lambda_1^2  \lambda_2^2 
   \lambda_4^2} \,, 
\quad 
M_2 =
\frac{\beta_t^2}{\lambda_4^2}-\frac{2}{\lambda_3^2} \,, 
   \\
 M_{3,4}  &=& 
   \frac{   \beta_t^2}{\lambda_4^2}-\frac{2}{\lambda_3^2} \pm 
   \frac{\sqrt{   \lambda_3^4 
   \left(\lambda_1^2 -  \lambda_2^2 \right)^2
   \left(\left(\beta_t^2+2\right) 
   \lambda_1^2 \lambda_2^2 +\lambda_1^4+\lambda_2^4\right)}}{2 \lambda_1^2
   \lambda_2^2 \lambda_3^2 \lambda_4^2} \,. 
\eea
Similar to the $A_{4,10}$ case, we see that in this case $M_{1}$ is negative
and hence the NEC is violated, unless $\lambda_1=\lambda_2$. 

In the case $\lambda_1=\lambda_2$, 
the geometry becomes a product of the form 
\bea
ds^{2} &=& - e^{2\beta_{t}r}dt^{2}+ \lambda_4^2 {dr^{2}} + \lambda_3^2 d \vec{x^{2}}_{EAdS3} \,,
\eea
where $d \vec{x^{2}}_{EAdS3}$ is the Euclidean $AdS_3$ metric with unit length-scale. 
In this case the NEC then imposes just that $\beta_{t}^{2}>2 \lambda_{4}^{2}/\lambda_3^2$. 

{\it Case 2: $r  = x^3$}

If we pick $r=x^{3}$ instead, the eigenvalues of $M$ are then
\bea
M_1 &=&  -\frac{2 (\beta_t+1)}{\lambda_3^2} \,, \quad 
M_2 = \frac{(\beta_t-2)  \beta_t}{\lambda_3^2} - \frac{\left(\lambda_1^2
-\lambda_2^2\right)^2}{2 \lambda_1^2 \lambda_2^2 \lambda_4^2} \,, \\
M_{3,4} &=& \frac{(\beta_t-2)  (\beta_t+1)}{\lambda_3^2}
\pm \frac{\lambda_1^4-\lambda_2^4}{2
   \lambda_1^2 \lambda_2^2 \lambda_4^2} \,.
\eea
Nernst's law forces 
\bea
\beta_t < 0 \,.
\eea
Therefore, both the NEC and Nernst's law are easily satisfied at large negative $\beta_t$. 
\end{itemize}

In summary, we have seen that the types of space-times that are eliminated 
by the NEC are types $A_{4,1}$, $A_{4,8}$, $A_{4,10}$, and $A_{4,12}$ with the $x^4 =r $ choice.  
The rest of the classes all contain at least one open set of parameters
where both the NEC and ``Nernst's law" are satisfied, around large negative values of $\beta_t$. 

The reader should note that 
we have made the ``obvious" choice of the holographic radial coordinate 
in each case in the analysis above.\footnote{``Obvious'' since we have taken $r$ simply 
identified with one of the $x^i$, but generically we can take more 
complicated combinations of them as radial coordinate. For example, in the type $A_{4,12}$, 
one can also take any linear combinations of $x^3$ and $x^4$ as radial coordinates, $r$.}
Generically, the radial direction $r$ can be a more complicated function of $x^i$'s. 
While this freedom does not enter in any of the data involving the spatial 4-slices, 
it is relevant in the construction of our 5d space-time. 
This is because we have selected the redshift factor $g_{tt}(r)$ ``by hand" 
to have the form $e^{2 \beta_t r}$ and 
its contributions in the NEC calculations clearly depend upon, among other things, which coordinate we choose to consider as $r$. 
It is quite possible that more elaborate choices than those considered here could resurrect some
of the algebraic structures we have left for dead.

\section{5d space-time avatars of four-algebras}{\label{sec:5d4algebra}}

In this section we will demonstrate that some of the afore-mentioned geometries
are indeed realizable from reasonable matter content. In particular,
we will work out a few examples explicitly, in a system with a similar
effective action to one we've discussed before (eq.(\ref{EHAction_S_V_L})):\footnote{
In fact the above action is the same as eq.(\ref{EHAction_S_V_L}) generalized to include two
gauge fields, up to the following simple field redefinitions:
$\phi_{here} = \frac{1}{2} \phi_{there},~\alpha_{here} = 2 \alpha_{there},~ \delta_{here}=2 \delta_{there},~\beta =2 \epsilon$.}
\begin{equation}
S=\int dx^{5}\sqrt{-g}\left\{ R-2(\nabla\phi)^{2}+e^{2\delta\phi}\Lambda-\frac{1}{4}e^{2\alpha\phi}(F_A^2 + F_B^2) -\frac{1}{4}e^{2\beta\phi}(M_{A}^{2}A^{2} + M_B^2 B^2)\right\} ~
\end{equation}
with $A, B$ two abelian gauge fields, and $F_A, F_B$ their field strengths.

However, we are going to relax the requirement that the spacetime
be scale-invariant, and look for a possibly hyperscaling violating
generalization of the space times based on four-algebras. In the orthonormal basis notation, this
means that we are going to generalize the radial one-form to be $$\sigma^{r}=\frac{dr}{f(r)}$$
where $$f(r)=e^{\theta r}$$ for a hyperscaling violating metric (with hyperscaling-violation exponent $\theta$). The structure
constants involving the radial index will need to change via: $C_{rj}^{i}\rightarrow C_{rj}^{i}f(r)$.
The equations of motion will be modified in the orthonormal basis to
be:
\begin{itemize}
\item Electric field:
\[
\left\{ \left[A_{t}^{'}(r)f(r)+A_{t}(r)C_{rt}^{t}(r)\right]e^{2\alpha\phi}\right\} ^{'}f(r)+\left[A_{t}^{'}(r)f(r)+A_{t}(r)C_{rt}^{t}(r)\right]C_{ri}^{i}e^{2\alpha\phi}
\]
\begin{equation}
=\frac{M_A^{2}}{2}e^{2\beta\phi}A_{t}(r)
\end{equation}
with a similar equation for $B$.

\item Magnetic field:
\begin{equation}
\left[A_{i}^{'}(r)f(r)+A_{j}(r)C_{ri}^{j}(r)\right]C_{pq}^{k}\epsilon^{ikl}\epsilon^{pql}e^{2\alpha\phi}=0
\end{equation}
\[
\frac{M_A^{2}}{2}e^{2\beta\phi}A_{m}(r)=\left\{ \left[A_{m}^{'}(r)f(r)+A_{j}(r)C_{rm}^{j}(r)\right]e^{2\alpha\phi}\right\} ^{'}f(r)
\]
\[
+\left[A_{m}^{'}(r)f(r)+A_{j}(r)C_{rm}^{j}(r)\right]C_{rt}^{t}(r)e^{2\alpha\phi}
\]
\begin{equation}
-\frac{1}{4}A_{i}(r)C_{pq}^{i}C_{kl}^{j}\epsilon^{jpq}\epsilon^{mkl}e^{2\alpha\phi}+[A_{i}^{'}(r)f(r)+A_{j}(r)C_{ri}^{j}(r)]C_{rp}^{k}(r)\epsilon^{ikl}\epsilon^{mpl}e^{2\alpha\phi}
\end{equation}
with a similar equation for $B$.
\item Scalar field:
\[
2\phi'(r)f(r)\epsilon^{mjk}C_{ri}^{m}(r)\epsilon^{ijk}+4\phi'(r)f(r)C_{rt}^{t}(r)+4(\phi'(r)f(r))'f(r)
\]
\begin{equation}
-\frac{\alpha}{2}e^{2\alpha\phi}(F_{A}^{2} + F_B^2)-\frac{\beta}{2}e^{2\beta\phi}(M_{A}^{2}A^{2} + M_B^2 B^2) +2\delta e^{2\delta\phi}\Lambda=0
\end{equation}
$ $The gauge curvature is given by
\[
F_A(r)=\left[A_{i}^{'}(r)f(r)+A_{j}(r)C_{ri}^{j}(r)\right]\sigma^{r}\wedge\sigma^{i}
+[A_{t}^{'}(r)f(r)+A_{t}(r)C_{rt}^{t}(r)]\sigma^{r}\wedge\sigma^{t}
\]
\begin{equation}
+ \frac{1}{2}A_{i}(r)C_{jk}^{i}\sigma^{j}\wedge\sigma^{k}
\end{equation}
with a similar equation for B.
\end{itemize}

\subsection{$A_{4,2}^{a}$ }

Many of the space-times in the classification before are simply scaling
geometries with a spatial 3-fold belonging to one of the 3d Bianchi
types. They have already been dealt with in previous sections. We
will therefore focus on the geometries with non-trivial radial actions
on the spatial geometries, and $A_{4,2}^{a}$ is the first such case.
A hyperscaling violating version of the $A_{4,2}^{a}$ geometry takes
the following metric:$ $
\begin{equation}
ds^{2}=-e^{2\beta_{t}r}dt^{2}+\frac{dr^{2}}{f(r)^{2}\beta_{1}^{2}}+ ( \omega^{1})^{2}+\beta_{2}^{2} (\omega^{2})^{2}+ (\omega^{3})^{2}
\end{equation}

The structure coefficients in the orthnomal basis becomes:
\begin{equation}
C_{34}^{2}(r)=\beta_{2}\beta_{1}f(r)\,, \quad C_{14}^{1}(r)=a\beta_{1}f(r) \,, \quad C_{24}^{2}(r)=C_{34}^{3}(r)=\beta_{1}f(r) \,,
\end{equation}
where again
$f(r)=e^{\theta r}$. We will turn on the two massive vector fields and the dilaton as: 
\bea
A(r)=A_{t}e^{-\omega\phi(r)}\sigma^{t} \,, \quad 
B(r)=B_{2}e^{-\omega\phi(r)}\sigma^{2}+B_{3}e^{-\omega\phi(r)}\sigma^{3} \,,  \quad
%as well as the dilaton $
\phi(r)=k r \,. \nonumber
\eea 
If we take
\[
\alpha=\beta-\frac{\theta}{k} \,, \quad \delta=\frac{\theta}{k} \,, \quad \theta=k(\beta-\omega) \,,
\]
then the equations of motion becomes algebraic:
\begin{itemize}
\item EOM for A:
\begin{equation}
M_{A}^{2}+2(2+a-k\beta)\beta_{1}^{2}(\beta_{t}-k\omega)=0
\end{equation}

\item EOM for B:
\bea
2B_{3}\beta_{1}^{2}\beta_{2}(1+k\omega)+B_{2}(M_{B}^{2}+2\beta_{1}^{2}(-1-a+k\beta+\beta_{2}^{2}
%\]
\quad \quad \quad \quad \quad \quad \quad \quad \quad \quad  \quad   \nonumber \\
%\begin{equation}
+\beta_{t}+k(-1-a+k\beta+\beta_{t})\omega))= 0 \quad \\
%\end{equation}
%\begin{equation}
2B_{2}\beta_{1}^{2}\beta_{2}(-1-a+k\beta+\beta_{t})+B_{3}(M_{B}^{2}-2\beta_{1}^{2}(1+a-k\beta-\beta_{t})(1+k\omega))=0 \quad
%\end{equation}
\eea

\item EOM for $\phi$:
\bea
-B_{2}^{2}M_{B}^{2}\beta-B_{3}^{2}M_{B}^{2}\beta-8(2+a)k\beta_{1}^{2}+8k^{2}\beta\beta_{1}^{2}+8k\beta_{1}^{2}\beta_{t}+4\Lambda(\beta-\omega)
\quad \quad \quad \quad \quad \quad \nonumber \\
%\]
%\[
-2\beta_{1}^{2}(B_{3}^{2}+4k^{2}+B_{2}^{2}(1+\beta_{2}^{2})+2\beta_{2}B_{2}B_{3})\omega
-4k\beta_{1}^{2}(B_{2}^{2}+B_{3}^{2}+B_{2}B_{3}\beta_{2})\omega^{2}
%\]
%\begin{equation}
\quad \quad \quad \nonumber \\
-2(B_{2}^{2}+B_{3}^{2})k^{2}\beta_{1}^{2}\omega^{3}+A_{t}^{2}(M_{A}^{2}\beta+2\beta_{1}^{2}\omega(\beta_{t}-k\omega)^{2})=0 \quad \quad 
%\end{equation}
\eea

\item Einstein's Equations:
\bea
-4\Lambda+B_{3}^{2}(M_{B}^{2}+2\beta_{1}^{2})+4B_{2}B_{3}\beta_{1}^{2}\beta_{2}+B_{2}^{2}(M_{B}^{2}+2\beta_{1}^{2}(1+\beta_{2}^{2}))
%\]
%\[
\quad \quad \quad \quad \quad \quad  \quad \quad \quad  \nonumber \\
+2\beta_{1}^{2}(12+\beta_{2}^{2}-8\beta_{t}+4(k^{2}+k\beta(-2+\beta_{t})+\beta_{t}^{2}))
-A_{t}^{2}(M_{A}^{2}+2\beta_{1}^{2}(\beta_{t}-k\omega)^{2})
%\]
%\begin{equation}
\quad \quad \quad  \nonumber \\
+2k\beta_{1}^{2}\omega(8+2B_{2}B_{3}\beta_{2}-4\beta_{t}+B_{2}^{2}(2+k\omega)+B_{3}^{2}(2+k\omega))=0 \quad   \quad  
\eea
%\end{equation}
%\[
%\]%%%%%%%%%%%%%%%%%%%
%\[
\bea
&& -4\Lambda+B_{3}^{2}(M_{B}^{2}+2\beta_{1}^{2})+4B_{2}B_{3}\beta_{1}^{2}\beta_{2}-B_{2}^{2}(M_{B}^{2}-2\beta_{1}^{2}(-1+\beta_{2}^{2})) \quad \quad \quad \quad 
%%
%
%\]
%\[
  \nonumber \\
&& \quad +2\beta_{1}^{2}(4+4a^{2}+3\beta_{2}^{2}-4\beta_{t}+4(k^{2}+k\beta(-1+\beta_{t})+\beta_{t}^{2})-4a(-1+k\beta+\beta_{t}))
\quad \quad \nonumber \\
%\]
%\begin{equation}
&&\quad \quad \, \,  \, -A_{t}^{2}(M_{A}^{2}+2\beta_{1}^{2}(\beta_{t}-k\omega)^{2}) \nonumber \\
&& \quad \quad \quad \quad \, \,
+2k\beta_{1}^{2}\omega(4+4a+2B_{2}B_{3}\beta_{2}-4\beta_{t}-B_{2}^{2}(2+k\omega)+B_{3}^{2}(2+k\omega))=0 \quad \quad \quad 
%\end{equation}
\eea
%\[
%\]
\begin{equation}
-2B_{2}^{2}\beta_{1}^{2}\beta_{2}(1+k\omega)-2\beta_{1}^{2}\beta_{2}(2+a-k\beta-\beta_{t}+k\omega)-B_{2}B_{3}(M_{B}^{2}+2(\beta_{1}+k\beta_{1}\omega)^{2})=0
\end{equation}
%\[
%\]
%\[
\bea
&& -4\Lambda-B_{3}^{2}(M_{B}^{2}+2\beta_{1}^{2})-4B_{2}B_{3}\beta_{1}^{2}\beta_{2}+B_{2}^{2}(M_{B}^{2}-2\beta_{1}^{2}(-1+\beta_{2}^{2}))
%\]
\nonumber \\
%\[
&& \quad +2\beta_{1}^{2}(4+4a^{2}-\beta_{2}^{2}-4\beta_{t}+4(k^{2}+k\beta(-1+\beta_{t})+\beta_{t}^{2})-4a(-1+k\beta+\beta_{t}))
%\]
\nonumber \\
%\begin{equation}
&& \quad \quad -A_{t}^{2}(M_{A}^{2}+2\beta_{1}^{2}(\beta_{t}-k\omega)^{2})
\nonumber \\
&& \quad \quad \quad
+2k\beta_{1}^{2}\omega(4+4a-2B_{2}B_{3}\beta_{2}-4\beta_{t}+B_{2}^{2}(2+k\omega)-B_{3}^{2}(2+k\omega))=0
%\end{equation}
\eea
%\[
%%%%%%%%%%%%%%%%%%%%%%%
%\]
\bea
%\[
&& -4\Lambda+B_{3}^{2}(M_{B}^{2}-2\beta_{1}^{2})-4B_{2}B_{3}\beta_{1}^{2}\beta_{2}+B_{2}^{2}(M_{B}^{2}-2\beta_{1}^{2}(1+\beta_{2}^{2}))
\nonumber \\
&& \quad \quad -2\beta_{1}^{2}(-4+4k^{2}+\beta_{2}^{2}+4a(\beta_{t}-2)+8\beta_{t})
%\]
%\begin{equation}
%\nonumber \\
-A_{t}^{2}(M_{A}^{2}
-2\beta_{1}^{2}(\beta_{t}-k\omega)^{2})  
\nonumber \\
&& \quad \quad  \quad \quad 
-2k\beta_{1}^{2}\omega(2B_{2}B_{3}\beta_{2}+B_{2}^{2}(2+k\omega)+B_{3}^{2}(2+k\omega))=0
%\end{equation}
\eea
%\[
%%%%%%%%%%%%%%%%%%%%%%%
%\]
%\[
\bea
&& 4\Lambda-B_{3}^{2}(M_{B}^{2}+2\beta_{1}^{2})-4B_{2}B_{3}\beta_{1}^{2}\beta_{2}
-B_{2}^{2}(M_{B}^{2}+2\beta_{1}^{2}(1+\beta_{2}^{2}))
\nonumber \\
&& \quad  -2\beta_{1}^{2}(4(3+2a+a^{2}+k^{2}-(a+2)k\beta)+\beta_{2}^{2})
%\]
%\begin{equation}
-A_{t}^{2}(M_{A}^{2}+2\beta_{1}^{2}(\beta_{t}-k\omega)^{2}) 
\nonumber \\
&& \quad \quad -2k\beta_{1}^{2}\omega(8+4a+2B_{2}B_{3}\beta_{2}+B_{2}^{2}(2+k\omega)+B_{3}^{2}(2+k\omega))=0
%\end{equation}
\eea
A general solution is impractical for such a system of algebraic equations.
Instead, we will show that a solution exists for some particular values
of the parameters, which gives a thermodynamically reasonable space-time.
In particular, we pick $a=2,\beta_{1}=\frac{3}{2},\beta_{t}=-2,\beta_{2}=2$.
The following solution is obtained:
\[
k=-\frac{6}{\epsilon} \,, \quad \Lambda=-54+\frac{162}{\epsilon^{2}}\,, \quad 
B_{2}=\epsilon\frac{\sqrt{2(1+\sqrt{26+\frac{12\beta(3\beta+5\epsilon)}{\epsilon^{2}}})}}{6\beta+5\epsilon} \,, 
\]
\[
A_{t}=\frac{\sqrt{36+\epsilon^{2}(-12+\sqrt{26+\frac{12\beta(3\beta+5\epsilon)}{\epsilon^{2}}})}}{3\beta+2\epsilon} \,, 
\]
\[
B_{3}=B_{2}\frac{\epsilon(1+\sqrt{26+\frac{12\beta(3\beta+5\epsilon)}{\epsilon^{2}}})}{6\beta+5\epsilon}\,, \quad M_{A}^{2}=-18\frac{(3\beta+2\epsilon)^{2}}{\epsilon^{2}} \,, 
\]
\begin{equation}
M_{B}^{2}=-\frac{9}{2\epsilon^{2}}(36\beta^{2}+60\beta\epsilon+\epsilon^{2}(27-2\sqrt{26+\frac{12\beta(3\beta+5\epsilon)}{\epsilon^{2}}})) \,,
\end{equation}
where $\epsilon=\omega-\beta$. We see that a large open set of $\beta,\epsilon$
will yield real solutions with a positive value of $\Lambda$ (so
that it is, quite plausibly, ultimately gluable to $AdS_{5}$).  For example, taking $\beta=-{1\over 2}$, $\epsilon=1,$
we have $k=-6, A_{t}=2 \sqrt{24 + \sqrt{5}}, B_{2}=\sqrt{(1 + \sqrt{5})/2},B_{3}=( 1 + \sqrt{5})^{3/2}/ 2\sqrt{2}, \Lambda=108,M_{A}^{2}=-{9\over 2},M_{B}^{2}=-{9 \over 2} (6 - 2\sqrt{5})$.
Also notice that the solution is not valid when $\epsilon=0$, which
corresponds to the conformal case $\theta=0$. This fact remains true
for other values of the parameters, indicating that in this system $A_{4,2}^{a}$
is not realizable unless it has non-zero hyperscaling violation exponent $\theta$.
\end{itemize}

\subsection{$A_{4,3}$}

This is a simplified version of $A_{4,2}^{a}$ with the structure
coefficients
\begin{equation}
C_{34}^{2}(r)=\beta_{2}\beta_{1}f(r) \,, \quad C_{14}^{1}(r)=\beta_{1}f(r) \,.
\end{equation}

It can be shown that with the same matter content as the previous
example, this class of geometries is also realizable. Again, if we fix
that $\beta_{1}=\frac{3}{2},\beta_{t}=-2,\beta_{2}=2$, the solutions
take the form:
\[
k=-\frac{3}{\epsilon} \,, \quad 
\Lambda=-\frac{9}{2}+\frac{81}{2\epsilon^{2}} \,, \quad 
B_{2}=\epsilon\frac{\sqrt{2(1+\sqrt{10+\frac{9\beta(\beta+2\epsilon)}{\epsilon^{2}}})}}{3(\beta+\epsilon)} \,,
\]
\[
A_{t}=\frac{2\sqrt{9+\epsilon^{2}(-1+\sqrt{10+\frac{9\beta(\beta+2\epsilon)}{\epsilon^{2}}})}}{3\beta+\epsilon} \,, \quad M_{A}^{2}=-\frac{9(3\beta+\epsilon)^{2}}{2\epsilon^{2}} \,,
\]
\[
B_{3}=B_{2}\frac{\epsilon(1+\sqrt{10+\frac{9\beta(\beta+2\epsilon)}{\epsilon^{2}}})}{3(\beta+\epsilon)} \,, 
\]
\begin{equation}
M_{B}^{2}=-\frac{9}{2\epsilon^{2}}(9\beta^{2}+18\beta\epsilon+\epsilon^{2}(11-2\sqrt{10+\frac{9\beta(\beta+2\epsilon)}{\epsilon^{2}}})) \,.
\end{equation}

Similarly to the previous case, a large open set of $\beta$ and $\epsilon$ can yield real
solutions with positive $\Lambda$; and this system can only support geometries
with non-zero hyperscaling violation.

\subsection{$A_{4,6}^{a,b}$}

This geometry has the interesting feature that the radial action involves
both scaling and rotation on the spatial 3-manifold. The hyperscaling
violating version of the geometry has the following metric:
\[
ds^{2}=-e^{2\beta_{t}r}dt^{2}+\frac{dr^{2}}{f(r)^{2}\beta_{1}^{2}}+ (\omega^{1})^{2}+ (\omega^{2})^{2}+\lambda^{2} (\omega^{3})^{2}
\]
with structure coefficients: 
\[
C_{14}^{1}(r)=a\beta_{1}f(r) \,, \quad C_{24}^{2}(r)=C_{34}^{3}(r)=b\beta_{1}f(r) \,, \quad 
C_{34}^{2}(r)=\frac{\beta_{1}}{\lambda}f(r) \,, \quad C_{24}^{3}(r)=-\lambda\beta_{1}f(r)
\]

Since the radial action rotates the 2-3 plane of the geometry, choosing
one of the vector fields to be aligned with either of $\sigma^{2},\sigma^{3}$
will suffice to be general. Therefore we will turn on the following
fields:
\[
A(r)=A_{t}e^{-\omega\phi(r)}\sigma^{t} \,, \quad 
B(r)=B_{2}e^{-\omega\phi(r)}\sigma^{2} \,, \quad \phi(r)=kr
\]

As before, taking $\alpha=\beta-\frac{\theta}{k},\delta=\frac{\theta}{k},\theta=k(\beta-\omega)$
will reduce the equations to algebraic equations:
\begin{itemize}
\item $A_{t}$:
\begin{equation}
M_{A}^{2}+2(a+2b-k\beta)\beta_{1}^{2}(\beta_{t}-k\omega)=0
\end{equation}

\item $B_{2}:$
%\begin{equation}
\bea
M_{B}^{2}\lambda^{2}+2\beta_{1}^{2}(1-(a+b-k\beta-\beta_{t})\lambda^{2}(b+k\omega))=0 &&
%\end{equation}
\\
%\begin{equation}
a+b-k\beta-\beta_{t}+b\lambda^{2}+k\lambda^{2}\omega=0 &&
%\end{equation}
\eea

\item $\phi:$ 
\bea
(B_{2}^{2}M_{B}^{2}\beta-8k\beta_{1}^{2}(-a-2b+k\beta+\beta_{t}))\lambda^{2}+2\beta_{1}^{2}(B_{2}^{2}+(b^{2}B_{2}^{2}+4k^{2})\lambda^{2})\omega
%\]
\quad \quad \quad \nonumber \\
%\[
+4bB_{2}^{2}k\beta_{1}^{2}\lambda^{2}\omega^{2}+2B_{2}^{2}k^{2}\beta_{1}^{2}\lambda^{2}\omega^{3}+4\Lambda\lambda^{2}(-\beta+\omega)
%\]
\quad \quad  \quad   \nonumber \\
%\begin{equation}
-A_{t}^{2}\lambda^{2}(M_{A}^{2}\beta+2\beta_{1}^{2}\omega(\beta_{t}-k\omega)^{2})=0 \quad \quad
%\end{equation}
\eea

\item Einstein's Equations:
\bea
(-4\Lambda-A_{t}^{2}M_{A}^{2}+B_{2}^{2}M_{B}^{2})\lambda^{2}
+2\beta_{1}^{2} \Bigl(1+B_{2}^{2}(1+\lambda^{2}(b+k\omega)^{2})
%\]
\quad \quad \quad \nonumber \\
%\[
+\lambda^{2}(-2+12b^{2}-(-4+A_{t}^{2})\beta_{t}^{2}+\lambda^{2}-8b(\beta_{t}+k(\beta-\omega))
%\]
%\begin{equation}
\quad \quad \quad \nonumber \\
+2k\beta_{t}(2\beta+(A_{t}^{2}-2)\omega)+k^{2}(4-A_{t}^{2}\omega^{2})) \Bigr)=0
%\end{equation}
\eea
%\[
%\]
%\[
\bea
-(4\Lambda+A_{t}^{2}M_{A}^{2}+B_{2}^{2}M_{B}^{2})\lambda^{2}
+ 2\beta_{1}^{2} \Bigl(3+(-2+4(a^{2}+ab+b^{2}))\lambda^{2}
%\]
\quad \quad \quad \quad \quad \quad \quad \nonumber \\
%\[
-B_{2}^{2}((b\lambda+k\lambda\omega)^{2}-1)-\lambda^{2}(4k\beta(a+b-\beta_{t})+\beta_{t}(4(a+b)+(A_{t}^{2}-4)\beta_{t})
%\]
%\begin{equation}
\quad \quad \quad \nonumber \\
+\lambda^{2}-2k(2(a+b)+(A_{t}^{2}-2)\beta_{t})\omega+k^{2}(A_{t}^{2}\omega^{2}-4)) \Bigr)=0 
\quad \quad 
%\end{equation}
\eea
%\[
%\]
%\[
\bea
(-4\Lambda-A_{t}^{2}M_{A}^{2}+B_{2}^{2}M_{B}^{2})\lambda^{2}+2\beta_{1}^{2} \Bigl(-1+(-2+4(a^{2}+ab+b^{2}))\lambda^{2}
%\]
\quad \quad \quad \quad \quad \quad \nonumber \\
%\[
+B_{2}^{2}((b\lambda+k\lambda\omega)^{2}-1)+\lambda^{2}(-4k\beta(a+b-\beta_{t})-\beta_{t}(4(a+b)
%\]
\quad \quad  \quad \quad \quad \quad \quad \nonumber \\
%\begin{equation}
+(A_{t}^{2}-4)\beta_{t})+3\lambda^{2}+2k(2(a+b)+(A_{t}^{2}-2)\beta_{t})\omega+k^{2}(4-A_{t}^{2}\omega^{2})) \Bigr)=0 \quad 
%\end{equation}
\eea
%\[
%\]
%\[
\bea
(-4\Lambda-A_{t}^{2}M_{A}^{2}+B_{2}^{2}M_{B}^{2})\lambda^{2}
+2\beta_{1}^{2} \Bigl(-1+(2+4b^{2}-4k^{2}+a(8b-4\beta_{t})
%\]
\nonumber \\
%\begin{equation}
-8b\beta_{t}-\lambda^{2}+A_{t}^{2}(\beta_{t}-k\omega)^{2})\lambda^{2}-B_{2}^{2}(1+\lambda^{2}(b+k\omega)^{2}) \Bigr)=0
%\end{equation}
\eea
%\[
%\]
%\[
\bea
(4\Lambda-A_{t}^{2}M_{A}^{2}-B_{2}^{2}M_{B}^{2})\lambda^{2}
+2\beta_{1}^{2} \Bigl(-1-B_{2}^{2}((b\lambda+k\lambda\omega)^{2}+1)
\quad \quad 
\nonumber \\
-\lambda^{2}(-2+4a^{2}
%\]
%\begin{equation}
+12b^{2}+4k^{2}+A_{t}^{2}\beta_{t}^{2}+\lambda^{2}
%\nonumber \\
+8bk(\omega-\beta) \quad \quad 
\nonumber \\
+A_{t}^{2}k\omega(k\omega-2\beta_{t})+4a(2b+k(\omega-\beta))) \Bigr)=0
%\end{equation}
\eea
%\[
%\]
\begin{equation}
-b(2+B_{2}^{2}-2\lambda^{2})+a(\lambda^{2}-1)-(k\beta+\beta_{t})(\lambda^{2}-1)+k(\lambda^{2}-B_{2}^{2}-1)\omega=0
\end{equation}
%\[
%\]
Again, we are not going to give a general solution, but instead simply observe
that at $b=1,\beta_{t}=-2,\lambda=2$, the following solution is obtained:
%\[
%\]
\[
k=-\frac{1}{\omega} \,, \quad B_{2}=\sqrt{15} \,, \quad 
A_{t}=\frac{\sqrt{4+16\beta\omega+34\omega^{2}}}{\omega}
\]
\[
\Lambda=\frac{\beta_{1}^{2}(16+64\beta\omega+121\omega^{2})}{8\omega^{2}} \,, \quad 
a=-3-\frac{\beta}{\omega} \,,
\]
\begin{equation}
M_{A}^{2}=-2\beta_{1}^{2} \,, \quad M_{B}^{2}=-\frac{\beta_{1}^{2}}{2} \,.
\end{equation}
It is easy to see that by making for example $\beta=-4\omega\sim\varepsilon$
for some $\varepsilon\ll1$, we can obtain real solutions with $a>0,$
which appear thermodynamically stable; and $\Lambda>0$, which allows
it to be glued to $AdS_{5}.$ The conformal case $\theta=0\rightarrow\beta=\omega$
is apparently valid, except that it will make $a+2b<0$, hence is
not thermodynamically allowed (by ``Nernst's law" of \S8).
\end{itemize}

\subsection{Comment}

In summary, in this section, we have found (hyperscaling-violating analogues of) extremal
scaling metrics governed by several of the four-algebras consistent with the NEC and ``Nernst's
law."  It is interesting that we have been unable, as yet, to find such metrics without hyperscaling violationi, i.e. with 
 $\theta = 0$.
We leave this to further work. All the examples we have considered have a three-dimensional sub-algebra which acts on the 
spatial directions in which the field theory lives.
The gravity description suggests there could be even more novel possibilities where the bulk geometry is homogeneous with a 
symmetry group which does not have such a three-dimensional sub-algebra that acts on the field theory directions alone. 
An exploration of such geometries and their field theory duals is also left for the future.

%%%%%%%%%%%%%%%%%%%%%%%%%%%%%%%%%%%%%%%%%%%%%

%%%%%%%%%%%%%%%%%%%%%%%%%%%%%%%%%

\section{4d metrics governed by
three and four-algebras
}{\label{sec:Bianchi4dfor4d}}

In \S \ref{sec:NECfor5d} and \S \ref{sec:5d4algebra}, we have discussed 5d space-time. 
In this section, we discuss 4d space-time where the radial direction is nontrivially involved in 
realizations of the real three and four-algebras we've been discussing. 
We have seen that the fairly general static homogeneous horizons in 5d can 
be governed by a four-algebra 
involving the three ``field theory" spatial coordinates and the radial direction.   
Quite analogously, the Bianchi three-algebras can arise as symmetry algebras of
general horizons in 4d.  
In such cases, 
the Bianchi three-algebra mixes the two field theory spatial coordinates and the radial direction. 
In the first part of this section, we investigate examples where 
the radial and field theory spatial coordinates are intertwined in a non-trivial way 
by the Bianchi three-algebras.  Then, in the second part of this section, we investigate the 
examples where four-algebras can arise as symmetry algebras of horizons in 4d, 
where the radial and field theory spatial and time coordinates can all be intertwined.  
We restrict our attention to only the static metrics, and discuss conditions that the NEC
places on them. 
We will see that some of the types are excluded. 

\subsection{Bianchi three-algebras 
for radial and two-spatial directions}
We first discuss 4d space-times where the radial direction is nontrivially involved in the three-algebras.  
Here, we give some examples of this sort, which should be easily generalizable.

We will illustrate the type III examples.  
The Killing vectors $e_i$ of type III satisfy $\left[e_i,e_j\right] = C^k_{ij} e_k$ where $C^1_{13}=-C^1_{31}=1$ and the rest of the  $C^{i}_{j,k}=0$.
The Killing vectors and invariant one forms are given as\\
\begin{tabular}{cc}
$e_1=\partial_2$  & $\omega^1=e^{-x^1}dx^2$  \\
$e_2=\partial_3$  & $\omega^2=dx^3$\\
$e_3=\partial_1+x^2 \partial_2$  & $\omega^3=dx^1$ 
\end{tabular}\\

This three-algebra has three two-dimensional sub-algebras 
given by $\lbrace e_1,e_2 \rbrace$, $\lbrace e_2,e_3 \rbrace$ and $\lbrace e_1,e_3 \rbrace$.
We consider two possible embeddings of the two-algebras in the three-algebra.
\begin{itemize}

\item {\bf Case 1:} \\
Consider $\lbrace e_1,e_2 \rbrace$ to correspond to symmetries along the field theory spatial directions and $x^1=r$ to be the radial direction. 
Then the metric can be written as
\begin{equation}
\label{metric1}
ds^2= L^2 \left[ -e^{2 \beta_t r} dt^2+ dr^2 + e^{-2 r} (dx^2)^2+ (dx^3)^2 \right]~.
\end{equation}
Time is added as an extra direction, modifying $e_3$ to $e_3=\partial_r+x^2 \partial_2+\beta_t t \partial_t$
and adding $e_4=\partial_t$ to the set of Killing vectors.

\item {\bf Case 2:} \\
Consider $\lbrace e_1,e_3 \rbrace$ to correspond to symmetries along the field theory spatial directions and $x^3=r$ to be the radial direction. 
Then the metric can be written as
\begin{equation}\label{metric2}
ds^2= L_1^2 \left[-e^{2 \beta_t r} dt^2+ dr^2 \right]+L_2^2 \left[(dx^1)^2+ e^{-2 x^1} (dx^2)^2 \right]~.
\end{equation}
Time is added as an extra direction, modifying $e_2$ to $e_2=\partial_r+\beta_t t \partial_t$
and adding $e_4=\partial_t$ to the set of Killing vectors.
\end{itemize}

We will illustrate that both case 1 and case 2 can be obtained as solutions of 
Einstein gravity coupled to a (either massive or massless) gauge field with/without scalar fields.

\begin{itemize}
{\item \bf Case 1}

Case 1 can be obtained from the Einstein action coupled to a massive vector and a massless scalar field,
\begin{equation}
 S=\int dx^4 \sqrt{-g} ~\left[R+\Lambda-\frac{1}{4} F^2 -\frac{1}{4} m^2 A^2-2 (\partial \phi)^2\right]~.
\end{equation}
Along with the metric ansatz given by eq.(\ref{metric1}), we consider the ansatz for the gauge field and scalar field:
\begin{eqnarray}
 A &=& \sqrt{A_t} e^{\beta_t r} dt \,, \\
 \phi &=& \phi_1 x^3 \,.
\end{eqnarray}
The scalar field equation is identically satisfied. The vector field equation gives
\begin{equation}
 \sqrt{A_t} (m^2 L^2+2 \beta_t)=0~.
\end{equation}
The Einstein equations are given by
\begin{eqnarray}
%tt ~\text{ component:}~ 
 \frac{A_t}{L^2} \left( m^2 L^2 +2 \beta_t^2 \right) +8 (1+\phi_1^2)-4 \Lambda L^2&=&0 \nonumber \\
%rr~\text{ component:}~ 
 \frac{A_t}{L^2} \left( m^2 L^2 -2 \beta_t^2 \right) +8 (\beta_t-\phi_1^2)+4 \Lambda L^2&=&0 \nonumber \\
%x^2 x^2~\text{ component:}~ 
 \frac{A_t}{L^2} \left( m^2 L^2 +2 \beta_t^2 \right) -8 (\beta_t^2+\phi_1^2)+4 \Lambda L^2&=&0 \nonumber \\
%x^3 x^3~\text{ component:}~ 
 \frac{A_t}{L^2} \left( m^2 L^2 +2 \beta_t^2 \right) +8 (\beta_t-\beta_t^2-1+\phi_1^2)+4 \Lambda L^2&=&0 \nonumber \,.
\end{eqnarray}
The equations can be solved to get
\begin{eqnarray}
 m^2 L^2 &=& -2 \beta_t  \,, \\
\Lambda L^2 &=& 2-\beta_t +\beta_t^2 \,, \\
A_t &=& 2 L^2 (1+\frac{1}{\beta_t}) \,, \\
\phi_1^2 &=& \frac{1}{2}(1-\beta_t) \,.
\end{eqnarray}
So we have $\Lambda > 0$. 
In order to satisfy Nernst's law (i.e. that the horizon area vanishes at the horizon), 
we need $\beta_t < 0$, and therefore $m^2>0$ and $\phi_1^2 > 0$.
With this, $A_t>0$ implies $\beta_t<-1$.

{\item {\bf Case 2}}

Case 2 can be obtained as a solution of the Einstein-Maxwell action,
\begin{equation}
 S=\int dx^4 \sqrt{-g} ~\left[R+\Lambda-\frac{1}{4} F^2 \right]~.
\end{equation}
Along with the metric ansatz given by eq.(\ref{metric2}), we consider the gauge field ansatz to be
\begin{equation}
 A= \sqrt{A_t} e^{\beta_t r} dt~.
\end{equation}
The Maxwell equation is identically satisfied. 
Then the Einstein equations  
give only two independent equations:
\begin{eqnarray}
 \frac{A_t \beta_t^2}{L_1^2}+4 \frac{L_1^2}{L_2^2}-2 \Lambda L_1^2 &=& 0 \\
\frac{A_t \beta_t^2}{L_1^2}-4 \beta_t^2 +2 \Lambda L_1^2 &=& 0 ~.
\end{eqnarray}
The solution is given by
\begin{eqnarray}
A_t & = & 2 L_1^2 (1 + \frac{1}{1 - L_2^2 \Lambda}) \,,\\
\beta_t^2 &=& L_1^2 (-\frac{1}{L_2^2} + \Lambda) \,.
\end{eqnarray}
\end{itemize}
$\beta_t^2 > 0$ implies $\Lambda L_2^2 > 1$. 

%%%%%%%%%%%%%%%%%%%%%%%%%%%%%%%%%%%%%%%%%%%%%%

\subsection{Bianchi four-algebras $A_{4,k}$ in 4d Bianchi attractors}  

So far we investigated examples of the three-algebras realized in 4d space-time. 
We now investigate 4d realizations of the four-algebras $A_{4,k}$. 
Since this involves the field theory time coordinates as well, 
generically this induces a time-dependent metric. 
In this paper, we are rather interested in time-independent metrics, which may be dual
to (time-independent) ground states of doped field theories.  
Therefore we seek metrics which do not involve the time-coordinate explicitly. 
This leaves us with only two possibilities; either the metric should be static, or it should
be stationary. 
Here we restrict our attention to static metrics in 4d,  
but it is straightforward to generalize the analysis to include the stationary cases
\footnote{Once we allow the stationary cases, we have to worry about the possible presence of closed time-like curves, as seen in \cite{Bianchi}.}.

A suitable static metric is as follows. 
For simplicity, we consider the diagonal metric ansatz and takes the ``obvious'' choice 
for the coordinate identification.  
Then, the metric ansatz is 
\bea
\label{diagonalmetricansatz}
ds^2 = \sum_{i=1}^4 \eta_i (\omega^i )^2 \,
\eea 
where $\omega^i$ are the invariant one-forms of $A_{4,k}$ and we require all $\eta_i > 0$.  
Since the static property restricts the metric to the form
\bea
\label{4Dnontrivialtimemetric}
ds^2 = - e^{2 \beta_t r} dt^2 + ds^2_{3D \, Bianchi} \quad \,, 
\eea
we will see later that this form can be obtained only 
from the four-algebras $A^a_{4,2}$, $A_{4,3}$, $A^{a,b}_{4,5}$, $A^{a,b}_{4,6}$. 
The three dimensional subgroup $G$, which acts on the field theory spatial coordinates and radial direction, turns out to be type IV,  II, VI,  VII$_b$  in the 3d Bianchi's classification   
(or, $A_{3,2}$, $A_{3,1}$, $A_{3,5}^a$, $A_{3,7}^b$ in the notation of \cite{Patera})
for the $A_{4,k}$ with $k=2, 3, 5, 6$ respectively.

We will investigate the metric ansatz for each $A_{4,k}$ type given by eq.(\ref{diagonalmetricansatz}) and also
the corresponding null energy condition eq.(\ref{nullEconditionforarbitraryNN}) and 
``Nernst law" constraint as in \S8 
for each class of static metrics.  We choose our null vectors in investigating the NEC constraints
as in eq.(\ref{generalnullvector}), but here $i$ only runs over three values.

\begin{itemize}
\item{$A_{4,1}$:}

The invariant one-forms are manifestly dependent on $x^4$. 
So we should not choose $x^4$ as time if we wish to obtain a static metric.  
Furthermore, if we choose time as one of the $x^i$ ($i=1,2,3$),  
then the diagonal metric ansatz yields a stationary metric . 
For example, if we choose $x^1$ as the time and $x^4$ as the radial directions, the metric 
ansatz becomes like;
\bea
ds^2 = - (dt - r dx + \frac{1}{2} r^2 dy)^2 + \eta_r dr^2 +\eta_x (dx-r dy)^2 + \eta_y dy^2  \,,
\eea
and this generically induces closed time-like curves (CTC).  So we discard this case
without further exploration.

\item {$A^a_{4,2}$:}

Again in this case, we should not choose $x^4$ as time since then the 
metric will be manifestly time-dependent. 
Then, the best choice is 
to select $x^1$ as time and $x^4$ as the radial direction, yielding a metric ansatz 
\bea
\label{A42metric}
ds^2 = - e^{-2 a r }dt^2 + \eta_r dr^2 + e^{-2r} \left(\eta_x \left( dx - r dy \right)^2 + \eta_y dy^2 \right) \,,
\eea
which is static. We have $\eta_r > 0$, $\eta_x > 0$, $\eta_y > 0$. 
In this case, 
the spatial part of the metric coordinatized by ($x$, $y$, $r$) forms type IV of Bianchi's 3d 
classification, or $A_{3,2}$ in the notation of \cite{Patera}. 
(See Appendix \ref{Appendix:PateraBianchiDict}).

The non-zero components of the Einstein tensor are given by
\bea
&& G_{11} = -\frac{\eta_x + 12  \eta_y}{4 \eta_r \eta_y} \quad \,, \quad 
G_{22} =2 a + 1  - \frac{\eta_x}{4 \eta_y}  
\quad \,, \quad \\ 
&& G_{33} = \frac{\eta_x \left( 3 \eta_x + 4 (a^2 + a   + 1) \eta_y \right)}{4 \eta_r \eta_y}  \quad \,, \quad 
G_{34} = - \frac{(a + 2 )\eta_x}{2 \eta_r} \quad \,, \quad \\
&& G_{44} = -\frac{\eta_x - 4     (a^2 + a   + 1)       \eta_y}{4 \eta_r} \,.
\eea 
Then, for the arbitrary null vector  
$N^\mu = (\sqrt{{\sum_{i=1}^3 (s^i)^2}} \,, \frac{s^1}{\sqrt{\eta_r}} \,, \frac{s^2}{\sqrt{\eta_x}}   \,, \frac{s^3}{\sqrt{\eta_y}})$, 
the null energy condition gives:
\bea
N^\mu N^\nu T_{\mu\nu} 
= N^\mu N^\nu G_{\mu\nu}  
&=& \frac{f_1 (s^1)^2 + f_2 (s^2)^2 + f_3 s^2 s^3 + f_4 (s^3)^2 }{2 \eta_r \eta_y} \ge 0
 \,
\eea
where
\bea
f_1 &=& -\eta_x + 4 (a - 1)  \eta_y  \quad \,, \quad 
f_2 = \eta_x + 2 (a - 1) (a + 2 ) \eta_y \,, \\
f_3 &=& - 2 (a + 2 ) \sqrt{\eta_x \eta_y} \quad \,, \quad 
f_4 = -\eta_x  + 2 (a - 1) (a + 2 ) \eta_y  \,, 
\eea
for arbitrary $s^1$, $s^2$, and $s^3$. 
Therefore we need 
\bea
\label{nullECforA424D}
f_i \ge 0 \quad (i = 1, 2, 4) \quad \,, \quad 
4 f_2 f_4 \ge (f_3)^2  \,. \quad % (i = 1, 2, 4) \,.
\eea
Also, to satisfy Nernst's law, we need $a>0$. 

Let us set $\eta_x=1$ by a coordinate change. 
We can then show for large positive $a$ and $\eta_y$,
\bea
&& f_1 \to 4 a \eta_y > 0 \quad \,, \quad f_2 \to  2 a^2 \eta_y > 0  \quad \,, \quad
 f_4 \to 2 a^2 \eta_y > 0 \,, \nonumber \\
&& 4 f_2 f_4  \to 16 a^2 \eta_y^2 \gg (f_3)^2 \to  4 a^2 \eta_y   \,.
\eea
Therefore, eq.(\ref{nullECforA424D}) allows at least solutions at large ($a$, $\eta_y$). \\ \\
If we choose $x^2$ or $x^3$ as the radial direction instead, 
but keep using $x^1$ as time in order to obtain a static metric, 
the warp factor is not a function of the radial direction alone.
And if we do not choose time as $x^1$, then the metric becomes stationary at most, but not static. 
We postpone further investigation of these geometries.

\item {$A_{4,3}$:}

This has a very similar structure to the $A^a_{4,2}$ case above. 
So again to obtain a static metric, the simplest approach is to choose $x^1$ as time. 
If we choose $x^4$ as the radial direction, then the metric ends up with 
following form:  
\bea
\label{A43metric}
ds^2 = - e^{-2  r }dt^2 + \eta_r dr^2 +  \eta_x \left( dx - r dy \right)^2 + \eta_y dy^2  \,.
\eea
The only difference between this case and $A^a_{4,2}$ is 
the way in which radial warping appears in various components. 
In these cases, 
the spatial part of the metric coordinated by ($x$, $y$, $r$) form Bianchi's 3d 
classification type II, or $A_{3,1}$ %in the notation of \cite{Patera} 
(Appendix \ref{Appendix:PateraBianchiDict}). 

Note that in this case the horizon volume is independent of the radial coordinate. So special care must be taken for
analysis of Nernst's law or ``physical reasonableness".  Let us first check whether the Null Energy Condition is satisfied.\\ \\
The non-zero components of the Einstein tensor are given by
\bea
&& G_{11} = -\frac{\eta_x  }{4 \eta_r \eta_y} \quad \,, \quad 
G_{22} =  - \frac{\eta_x}{4 \eta_y} < 0 
\quad \,, \quad \\ 
&& G_{33} = \frac{\eta_x \left( 3 \eta_x + 4     \eta_y \right)}{4 \eta_r \eta_y}  \quad \,, \quad 
G_{34} = - \frac{\eta_x}{2 \eta_r} \quad \,, \quad \\
&& G_{44} = -\frac{\eta_x - 4          \eta_y}{4 \eta_r} \,.
\eea 
It follows that for a null vector 
$N^\mu = (1 \,, \frac{1}{\sqrt{\eta_r}} \,, 0\,, 0 )$,  
the Null Energy Condition is violated;
\bea
N^\mu N^\nu G_{\mu\nu}  
&=& - \frac{ \eta_x  }{2 \eta_r \eta_y} < 0 \,.
\eea
Therefore we cannot obtain this space-time from physical matter systems.

\item{$A_{4,4}$:}

This has a very similar structure to the $A_{4,1}$ case. 
Choosing $x^4$ as the radial direction, the warp factors as a function of r are the only difference. 
Again, there are three choices for time from $x^1$, $x^2$, $x^3$, but 
any choice always admits at most a stationary metric, but not a static one. 

\item{$A^{a,b}_{4,5}$:}

In this case, the simplest choice for a radial direction is $x^4$, since it admits warp factors in 
the metric which are
functions of $r$ only, consistent with the algebraic structure.  (If we do not choose $x^4$ as the radial direction, then the warp factors are not functions of radius alone). 
There are three equally good choice for time: $x^1$, $x^2$, or $x^3$. 
All give equivalently good static metric ansatzes. 
For example, setting $x^3$ to be time, we have a static metric 
\bea
\label{correspondstotypeIII}
ds^2 = - e^{- 2 b r}dt^2 + \eta_r dr^2  + e^{-2r} dx^2 + e^{- 2 a r} dy^2 \,.
\eea
Similarly setting either $x^1$ or $x^2$ as time, we obtain respectively, 
\bea
\label{A454dfornullEargument}
ds^2 = - e^{- 2 r}dt^2 + \eta_r dr^2  + e^{-2 a r} dx^2 + e^{- 2 b r} dy^2 \,,
\eea
and 
\bea
ds^2 = - e^{- 2 a r}dt^2 + \eta_r dr^2  + e^{-2r} dx^2 + e^{- 2 b r} dy^2 \,.
\eea
Note that these are just generalized Lifshitz geometries. 
In all cases, the three spatial coordinates ($x$, $y$, $r$) are in type VI of Bianchi's
classification, or $A_{3,5}^a$ (Appendix \ref{Appendix:PateraBianchiDict}) for generic $a \neq 0, 1$.  
For the special case $a=1$, the metric reduces to Bianchi's type V, or 
$A_{3,3}$ (Appendix \ref{Appendix:PateraBianchiDict}). \\ \\

Let's consider the metric ansatz  eq.(\ref{A454dfornullEargument}). It is pretty straightforward to calculate the Einstein tensor, and we obtain it in diagonal form as 
\bea
G_{11} &=& -\frac{a^2+b a+b^2}{\eta_r} \quad \,, \quad 
G_{22} =  b a+a+b \quad \,, \quad \\
G_{33} &=& \frac{b^2+b+1}{\eta_r}  \quad \,, \quad 
G_{44} = \frac{a^2+a+1}{\eta_r} \,.
\eea
Therefore, for the choice of null vectors
%\bea
$N^\mu = (1,\frac{1}{\sqrt{\eta_r}},0,0)$, % \quad \,, \quad 
$N^\mu = (1,0,1,0)$, % \quad \,, \quad 
$N^\mu = (1,0,0,1)$, % \,, 
%\eea 
respectively, we obtain 
\bea
\label{nv2andnv3}
 \frac{a - a^2 + b - b^2}{\eta_r}  \ge 0\,,  \quad  -\frac{(a-1) (a+b+1)}{\eta_r} \ge 0 \,, \quad  
-\frac{(b-1) (a+b+1)}{\eta_r} \ge 0 \,. \quad \quad
\eea

The black brane horizon is at $r \to \infty$ where $g_{tt} \to 0$. In order to 
satisfy the Nernst's law, we need 
\bea
e^{- (a + b )r } \to  0 \, \Rightarrow \, a + b > 0 \,.
\eea
Then, with $\eta_r > 0$, (\ref{nv2andnv3}) gives
\bea
a  \le 1 \quad \,, \quad b \le 1 \,.
\eea
Assuming that $g_{xx} \to 0$ and $g_{yy} \to 0$ at the horizon, 
we have a parameter regime where the Null Energy Condition is satisfied,  
\bea
0 \le a \le 1 \quad \,, \quad 0 \le b \le 1 \,.
\eea
Interestingly, on one boundary $a = b = 0$, we have $AdS_2 \times R^2$, 
and on the other boundary $a = b = 1$, we have $AdS_4$.

%%%%%%%%%%%%%%%%%%%%%%%%%%%%%%%%%%%%%%%%%%%%%%%

\item{$A^{a,b}_{4,6}$:}

Again, the best choice is to
select $x^1$ as time and $x^4$ as the radial direction. 
Then, we obtain the static metric ansatz 
\bea
\label{3dsubalgebramysterious}
ds^2 = e^{-2 a r} dt^2 + \eta_r dr^2 + e^{-2 b r} ( \eta_x \left(\cos r dx - \sin r dy \right)^2 + \eta_y \left( \cos r dy + \sin r dx \right)^2 ) \,.\nonumber \\
\eea
The three spatial coordinate ($x$, $y$, $r$) form Bianchi's 3d 
classification type  VII$_b$, or  $A^b_{3,7}$ in the 
notation of \cite{Patera}\footnote{The fact that the three coordinates ($x$, $y$, $r$) span a manifold of type VII$_b$ in Bianchi's 
classification, can be seen as follows (see \cite{Landau}).
We have chosen $e_1$ as the generator of time-translations, and the spatial subset $e_2, e_3, e_4$ 
forms a real three-algebra where antisymmetric structure constants are given by 
\bea
C^2_{24} =  b \, \,, \, C^3_{24} = -1\, \,,\, C^2_{34} = 1\, \,,\,  C^3_{34} = b \,. 
\eea
Defining the two-index constants $C^{dc}$ by $C^c_{ab} \equiv \epsilon_{abd} C^{dc}$, and 
furthermore decomposing them into a symmetric and anti-symmetric part as $C^{ab} \equiv n^{ab} + \epsilon^{abc} a_c$, we can see that $n^{ab}$ has two unit eigenvalues and one zero eigenvalue, and 
$a_c = (b, 0, 0)$. These are precisely the characteristics of Bianchi type VII$_b$.} (Appendix \ref{Appendix:PateraBianchiDict}).   \\ \\
It is pretty straightforward to calculate the Einstein tensor. 
Then, for the arbitrary null vector, $N^\mu = (\sqrt{{\sum_{i=1}^3 (s^i)^2}} \,, \frac{s^1}{\sqrt{\eta_r}} \,, \frac{s^2}{\sqrt{\eta_x}}   \,, \frac{s^3}{\sqrt{\eta_y}})$, 
we have the NEC 
\bea
N^\mu N^\nu T_{\mu\nu} 
= N^\mu N^\nu G_{\mu\nu}  
&=& \frac{f_1 (s^1)^2 + f_2 (s^2)^2 + f_3 s^2 s^3 + f_4 (s^3)^2 }{2 \eta_x \eta_y \eta_r} \ge 0
 \,
\eea
for arbitrary $s^1$, $s^2$, and $s^3$ with $\eta_r > 0$, $\eta_x > 0$, $\eta_y > 0$.
Here, the $f_i$ are given by:
\bea
\label{f1forA46typein4d}
f_1 &=&2 \eta_x \eta_y (2 b (a-b)+1)- \eta_x^2- \eta_y^2  \,, \\
f_2 &=& 2 \eta_x \eta_y (a-b) (a+2 b)+\eta_x^2-\eta_y^2 \,, \\
f_3 &=&  2 (a + 2 b) (- \eta_x + \eta_y)    \sqrt{\eta_x \eta_y} \,, \\ 
f_4 &=& 2 \eta_x \eta_y (a-b) (a+2 b)-\eta_x^2+\eta_y^2  \,. 
\label{f4forA46typein4d}
\eea

Therefore we need, 
\bea
\label{nullECforA464D}
f_i \ge 0 \quad (i = 1, 2, 4) \quad \,, \quad 
4 f_2 f_4 \ge (f_3)^2  \,. \quad % (i = 1, 2, 4) \,.
\eea
In Appendix \ref{Appendix:NEC} we discuss the explicit parameter ranges where these
conditions are satisfied.
\\

Other choices of coordinates yield either non-static metrics, or field-theory position-dependent warping factors in the metric.

\item{$A_{4,7}$ - $A_{4,12}$:}
By completely analogous reasoning to that appearing above, if we assume a diagonal metric ansatz and make the ``obvious'' 
choice(s) for the time and radial coordinates, we can obtain at most stationary metrics, but not static metrics. 

\end{itemize}

In summary, 
for the diagonal and static metric ansatz (\ref{diagonalmetricansatz}), the allowed cases for which the NEC and Nernst law constraints
can be met are,
 $A^a_{4,2}$,  $A^{a,b}_{4,5}$, and $A^{a,b}_{4,6}$.

%%%%%%%%%%%%%%%%%%%%%%%%%%%%%%%%%%%%%%%%%%% 

\section{Discussion}{\label{sec:discussion}}

In this paper, we have extended the program initiated in \cite{Bianchi}, 
of trying to classify a wide variety of less symmetric extremal near-horizon geometries 
for black branes, in several directions.  
We have demonstrated that one can simply modify these geometries to incorporate 
hyperscaling violation in the dual field theory.
We have given examples of such hyperscaling violation both in 4d and 5d bulk space-time metrics which realize the Bianchi types, with the radial direction included in the algebra. 
We have also shown that one can easily obtain analytical ``striped" metrics starting from the horizons
of \cite{Bianchi}.
It should be clear in each case that in providing examples, 
we have barely scratched the surface of what are likely very rich sets of solutions to the 
Einstein equations or appropriate low-energy limits of string theory.

In the direction of finding a more complete classification, 
we discussed the possible application of the larger algebraic structures uncovered in
\cite{Patera,MacCallum} 
to classify 5d extremal near-horizon geometries 
in terms of real four-algebras with preferred 3d subgroups.  
While we found that several
of these possibilities will remain unrealized in sensible gravity coupled to  matter theories 
(which satisfy the Null Energy Condition), others can be realized with simple
matter sectors coupled to gravity, and likely arise as duals to suitable infrared phases in
strongly-coupled quantum field theory.
Similarly, we have classified the 4d extremal static near-horizon space-times with 
symmetries governed by the
four-algebras, and we have seen that some of the types are forbidden due to the NEC,
but others are likely attainable.

A number of issues remain to be clarified.  In many cases, these near-horizon geometries manifest infrared singularities similar to that of the
Lifshitz space-time \cite{lifsol1,lifsings,lifsing}.  In the Lifshitz case, various physical smoothings or more subtle ``resolutions" of the metric singularity have been
discussed in \cite{HKW,Eva}.  The issue has also been addressed in isotropic space-times with hyperscaling violation in \cite{Cremonini,SandipB}.  
It would be interesting to see if similar physics arises for the anisotropic space-times described here.

In addition, while we have given evidence in \cite{Bianchi} that some of these horizons can be glued into asymptotically AdS space-time by suitable
RG flows, this has not been discussed in anything close to a comprehensive way for the full classes of anisotropic metrics we've described.  A careful study of which classes
of geometries are truly infrared phases of doped CFTs (perhaps with additional currents activated on the boundary) would be worthwhile.

Another important question is to study the stability of the  solutions found  both in this paper and in \cite{Bianchi}.
Such a study could include  analysing both   whether the solutions are perturbatively stable, i.e., whether they have modes 
which grow with time, and also whether they are stable with respect to changes in boundary conditions, i.e., the presence of relevant deformations with respect to RG flow. 

As always, one can wonder to what extent the rich set of possible phases found here manifest themselves in UV complete models derived from string theory.  It would be useful to 
explore embeddings of these solutions into gauged supergravity theories\footnote{See also \cite{Barisch} for the study of interesting black brane solutions in gauged supergravity.} that can be derived
from consistent truncation of IIA and IIB supergravity, for instance.  It would also be interesting to
find proposals for phases of matter in real systems which could give rise to some of the more exotic symmetry
groups discussed here.

A simple, possibly interesting, extension of this work is to relax the condition of the geometry being static. 
In \S10, we have seen that the real four-algebras in 4d space-time 
naturally induce metrics which are not static, but can be stationary. 
However, one must be careful with such metrics, since they can easily contain closed time-like curves,
as illustrated in \cite{Bianchi} (see also \cite{Oliver}). Classifying 4d stationary space-times governed by the four-algebras which
do not contain any such pathological features, could lead to interesting duals.
In  field theory systems,  
if external sources perturb the system, 
they  can induce currents to  flow.  This would naturally correspond to a stationary metric in the  putative 
gravity dual of the system. A concrete example  is that of a fluid subjected to a temperature or an electric potential which is 
time independent and varying slowly in the spatial directions. Gravity duals of extremal geometries 
subject to such potential
gradients would be worth studying further.\footnote{We thank Shiraz Minwalla, Mukund Rangamani and especially 
Sayantani Bhattacharyya for a discussion of this issue.}

The most ambitious possible extension would be to try and classify all 
 extremal inhomogeneous, anisotropic black brane horizons. 
In light of the interesting
scaling features shown in holographic transport in the simplest
inhomogeneous geometries \cite{Santos,SeanDonos}, this problem could be interesting for ``applied holography" in addition to its intrinsic interest as a
question in general relativity and string theory.
Needless to say, finding all such inhomogeneous phases  is a challenging question since the analysis 
cannot be reduced to merely solving  algebraic equations now and instead  requires  us to  
 confront coupled partial differential equations in their full glory. 
The striped phase discussed in \S6 is an example of  an inhomogeneous phase and our discussion  in that section
 can be viewed as a small step in this direction. Clearly though,  much more effort is needed to make progress on this issue.

%There has been some early work in this direction \cite{Haack,Banerjee,Mukund}, but the extremal limit (where one could hope to classify true ground states)
%manifests various subtleties that have yet to be resolved.  

%\newpage
\bigskip
\centerline{\bf{Acknowledgements}}
\medskip
We are grateful to N. Bao, S. Bhattacharyya,  K. Damle, S. Harrison, A. Karch, R. Loganayagam,  S. Minwalla, G. Mandal, 
 S. Nampuri, M. Rangamani, A. Sen,  T. Takayanagi and V. Tripathy  for very helpful discussions, and E. Fradkin, J. Gauntlett, S. Kivelson
and J. Maldacena for fun conversations after various talks about this material.
N.I. would like to thank the CERN COFUND fellowship, much of this work was done while 
he was a fellow at CERN. 
N.I. and S.P.T. would like to thank the organizers of 
a long-term workshop ``Gauge/Gravity Duality" held at YITP, Kyoto, for providing a stimulating atmosphere. 
S.K. would like to thank the organizers
of the ``Holographic Way" conference at Nordita, the theory groups at Imperial College and
Cambridge University, and the organizers of the ``Topology, Entanglement, and Strong Correlations in 
Condensed Matter" conference at the University of Illinois for providing stimulating environments as this work was completed.
N.K. would like to thank the organizers of the ``Advanced String School, 2012" held in Puri, India for their hospitality.
N.K., N.S.  and S.P.T. acknowledge funding from the Government of India, and thank the people of India for  
generously supporting research in string theory. S.P.T. thanks the ICTS and organisers of the workshop on ``Aspects of String Theory'', held in Bangalore.  
S.K. was supported in part by the US DOE under contract DE-AC02-76SF00515 and by the
National Science Foundation under grant no. PHY-0756174. P.N is grateful for the support of  Feinberg Postdoctoral Fellowship at the Weizmann Institute of Sciences where part of this work was  done.
 S.P.T. acknowledges support from a J. C. Bose Fellowship given by the 
Department of Science and Technology, Govt. of India. H.W is supported by a Stanford
Graduate Fellowship.

\newpage
\vfill
\begin{figure}
\begin{center}
\includegraphics{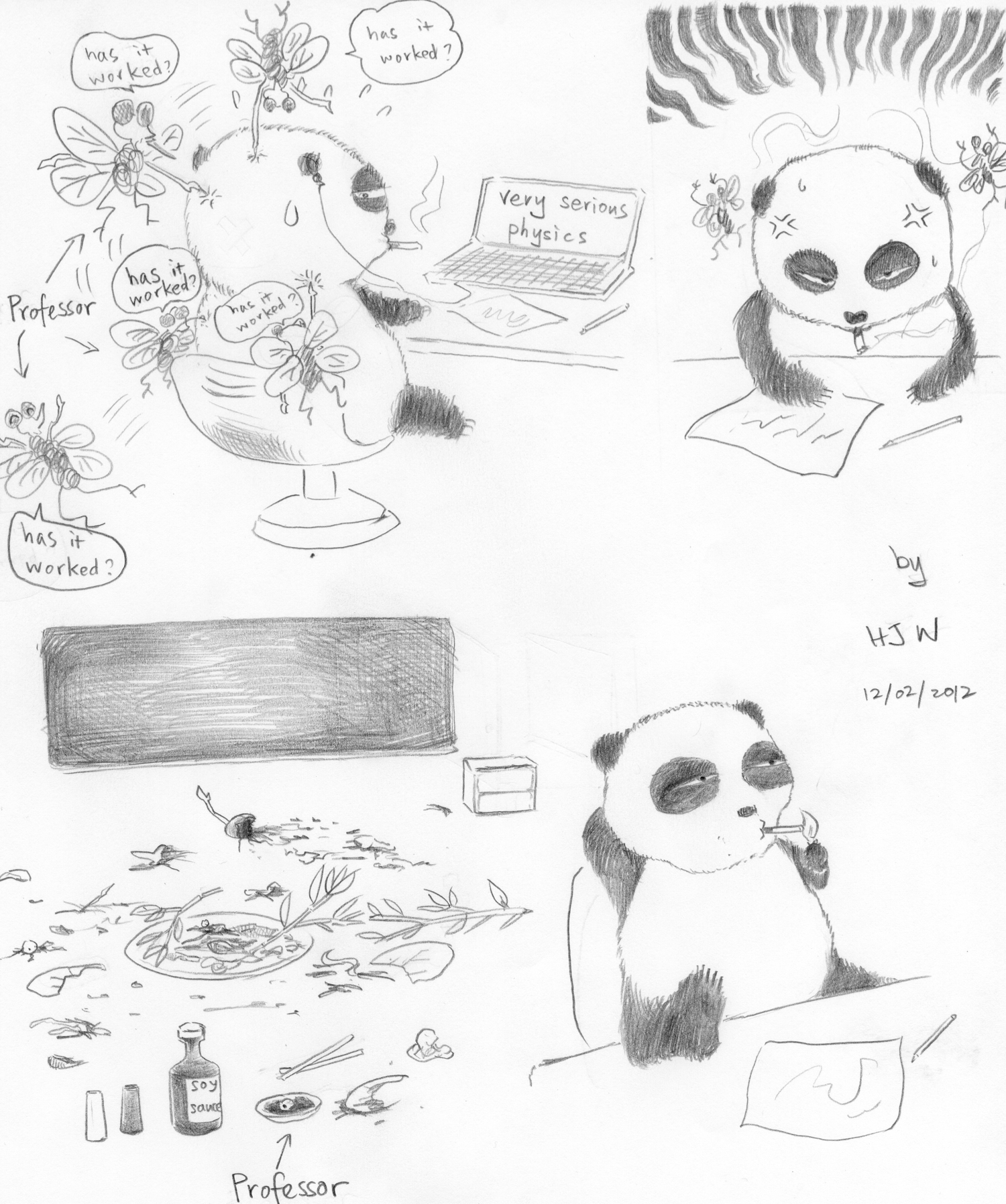}
\caption{Experience, and wishes, of a typical graduate student at Stanford or TIFR.}
\end{center}
\end{figure}
\vfill
\clearpage
%%%%%%%%%%%%%%%%%%%%%%%%%%%%%%%%%%%%%%%%%%%%%%%%%%%%%%%%%%%%
%%%%%%%%%%%%%%%%%%%%%%%%%%%%%%%%%%%%%%%%%%%%%%%%%%%%%%%%%%%%
\appendix

\section{Appendix A:  Einstein gravity coupled to massive vector fields} \label{Appendix:EOM}

In many sections of the paper, we will want to exhibit explicit solutions of Einstein
gravity coupled to a reasonable matter sector, to show that the near-horizon geometries
we propose can arise in physically sensible systems.   We will usually choose our
matter sector to consist of a set of massive Abelian gauge fields $A_a$, described
by the Lagrangian:

\begin{equation}
\label{ourmatter}
S=\int d^{5}x\sqrt{-g}\{R+\Lambda-\sum_{a}(\frac{1}{4}F_{a}^{2}+\frac{1}{4}m_{a}^{2}A_{a}^{2})\}
\end{equation}

In this section, we discuss some basic facts about the Einstein equations for the theory
with matter (\ref{ourmatter}).

\subsection{Ricci curvature}

We will be interested in static solutions of Einstein gravity.  The spatial slices spanned
by $\{ r,x_1, x_2, x_3 \}$ will
be homogeneous spaces as in \cite{Bianchi}, though we will allow for slightly more
general possibilities intertwining the radial and ``field theory" spatial coordinates in the
later sections of the paper.  

As in \cite{Bianchi}, the basic objects of interest are invariant one-forms $\omega^i$,
annihilated by the Killing vectors ${\mathrm e_i}$ which generate the 4-dimensional isometry group ${\cal G}$.  
Metrics which are written purely in terms of the $\omega^i$ with constant coefficients,
and with trivial $t$ dependence:

\begin{equation}
\label{trivmet}
ds^2 = -g_{tt}(r)dt^2 + \lambda_{ij} \omega^i \otimes \omega^j 
\end{equation}
will automatically be ${\cal G}$-invariant.  Here we slightly modify the notation of
\cite{Bianchi}; the factors of $e^{- \beta_i r}$ there, which incorporate the scaling of
the coordinates under radial translations, are built into the forms now, and the radial
scaling symmetry is treated as part of ${\cal G}$.

In order to exploit the underlying symmetries of the homogeneous geometries,
we will write down the Einstein-Maxwell equations in an
orthnomal (vielbein) basis, in which the metric tensor takes the form:
\bea
ds^{2}=\sum_{i=1}^{3} (\sigma^{i})^2 + (\sigma^{r})^2 - (\sigma^{t})^2
\eea
The vielbein elements are simple linear combinations of the $\omega$s; in
the most trivial case, $\sigma^i = \omega^i$.
In this formalism, we will in general lose
 the advantage that the one forms $\sigma^{\mu}$
are exact, and hence could be integrated into coordinates; instead
they form an orthonormal non-commuting basis. 

We are interested in scale-invariant (or conformally scale-invariant, in the
case of hyperscaling-violating metrics) near-horizon geometries. In order for the generalised scaling
to make sense, we require that the
radial coordinate $r$  to be identified with one of the exact one-forms $\sigma^r=\lambda_4 dr=dx^4$, 
such that the other 3 one-forms form a sub-algebra.
%In order for the generalised scaling
%to make sense, we require that one of the killing vectors, say $\mathrm{e_{4}}$, contains only $\partial_{4}$
%with no coefficient that depends on $\{x_{1},x_{2},x_{3}\}$, plus
%other derivative operators that generate the corresponding scaling
%transformations on the field theory spatial 3-manifold. 
%In requiring so, we can always
%write down a corresponding radial coordinate r, such that $\sigma^{r}=dr=dx_{4}$.
In this appendix A, we set $\lambda_4 =1 $. 
We further demand that $(\sigma^{t})^2 = g_{tt}(r)dt^{2}$, in keeping
with (\ref{trivmet}).  Therefore,
the orthnomal non-commuting basis satisfies the following commutation
relation:

\begin{equation}
\label{diffeq}
d\sigma^{\mu}=\frac{1}{2}C_{\nu\alpha}^{\mu}\sigma^{\nu}\wedge\sigma^{\alpha}~.
\end{equation}

The constants 
$C_{\alpha\beta}^{i}$ for $i\in\{1,2,3\}$ and $\alpha,\beta\in\{1,2,3,r\}$ are
given by the data of the four-algebra; $C_{\mu\nu}^{r}=0$ ; $C_{tr}^{t}=-\frac{1}{2 } g'_{tt}(r)/g_{tt}(r)$.
With this set-up the connection form $\Gamma_{\nu\beta}^{\mu}$ are
no longer given by the Christoffel symbols, but instead are given
by the structure constants via:

\begin{equation}
\Gamma_{\alpha\beta}^{\mu}=-\frac{1}{2}(g_{\tau\alpha}g^{\mu\sigma}C_{\sigma\beta}^{\tau}+g_{\tau\beta}g^{\mu\sigma}C_{\sigma\alpha}^{\tau}-C_{\alpha\beta}^{\mu})~.
\end{equation}
$g_{\mu\nu}$ is equal to the Minkowski $\eta_{\mu\nu}$ in this basis. See \cite{Landau} for detail. 
The Riemann curvature tensor is given in this basis by the connection
forms via:

\begin{equation}
R_{\mu\alpha\beta}^{\sigma}=\Gamma_{\mu\beta,\alpha}^{\sigma}-\Gamma_{\mu\alpha,\beta}^{\sigma}+\Gamma_{\mu\beta}^{\tau}\Gamma_{\tau\alpha}^{\sigma}-\Gamma_{\mu\alpha}^{\tau}\Gamma_{\tau\beta}^{\sigma}-(\Gamma_{\beta\alpha}^{\tau}-\Gamma_{\alpha\beta}^{\tau})\Gamma_{\mu\tau}^{\sigma}
\end{equation}
Hence we can compute the Ricci curvature $R_{\mu\nu}=R_{\mu\alpha\nu}^{\alpha}$
completely in terms of the structure constants, which only depend
on r. In the case where we also have scaling symmetry along the time
direction, the entire space-time is homogeneous and the Ricci curvature
is therefore algebraic.   We will see that in this formalism, the Einstein equations
reduce to algebraic equations, as in the story of generalized attractors discussed in 
e.g. \cite{Kallosh, Inbasekar:2012sh}.

\subsection{Maxwell's equations}

Here, we present the Maxwell equations for a single massive Abelian gauge field.  Since there are
no cross-couplings between the vector fields $A_a$, the generalization to multiple vectors is trivial.

Assume that the vector potential takes the form $A(r)=A_{t}(r)\sigma^{t}+\sum_{i}A_{i}(r)\sigma^{i}$.
Then the curvature is given by 

\begin{equation}
F=[A_{i}'(r)+A_{j}(r)C_{ri}^{j}]\sigma^{r}\wedge\sigma^{i}+[A_{t}'(r)+A_{t}(r)C_{rt}^{t}]\sigma^{r}\wedge\sigma^{t} + \frac{1}{2}A_{i}C_{jk}^{i}\sigma^{j}\wedge\sigma^{k}
\end{equation}
with components given by $F=\frac{1}{2}F_{\mu\nu}\sigma^{\mu}\wedge\sigma^{\nu}$.
The Maxwell equations for a massive vector field are given in differential
form as:

\begin{equation}
d\star_{5}F=-\frac{1}{2}m^{2}\star_{5}A
\end{equation}
Since the metric tensor is now Minkowskian, the Levi-Civita tensor reduces
to the usual flat case. It is therefore straight-forward to obtain the following
Maxwell's equations.

\subsubsection{Magnetic field}

In this case, $A_{t}(r)=0.$ The Maxwell's equations are:
\bea
%\begin{equation}
&& (A_{i}'(r)+A_{j}(r)C_{ri}^{j})C_{mn}^{k}\epsilon^{ikl}\epsilon^{mn l}=0 \quad \quad \quad \quad 
%\end{equation}
\eea
%\\
%
\bea
&& \frac{1}{2}m^{2}A_{m}(r)=(A_{m}'(r)+A_{j}(r)C_{rm}^{j})'+(A_{m}'(r)+A_{j}(r)C_{rm}^{j})C_{rt}^{t}
%\]
%
%
\quad \quad \nonumber \\
%\begin{equation}
&& \quad -\frac{1}{4}A_{i}(r)C_{pq}^{i}C_{kl}^{j}\epsilon^{j pq}\epsilon^{mkl}+(A_{i}'(r)+A(r)_{j}C_{ri}^{j})C_{r n}^{k}\epsilon^{ikl}\epsilon^{m n l}
%\end{equation}
\eea

\subsubsection{Electric field}

In this case, $A_{i}(r)=0.$ There is only one component of Maxwell's
equation:
\begin{equation}
(A_{t}'(r)+A_{t}(r)C_{rt}^{t})'+(A_{t}'(r)+A_{t}(r)C_{rt}^{t})C_{ri}^{i}=\frac{m^{2}}{2}A_{t}(r)
\end{equation}
Similar to before, in a scaling solution, we expect that the components
of the vector potential are constants, reducing Maxwell's
equations to a set of algebraic equations.

\subsection{Einstein's Equations}

The stress energy tensor of a massive gauge field is given by:

\begin{equation}
T_{\mu\nu}=\frac{1}{2}F_{\mu\lambda}F_{\nu}^{\lambda}+\frac{1}{4}m^{2}A_{\mu}A_{\nu}-\frac{1}{2}\eta_{\mu\nu}(\frac{1}{4}F_{\alpha\beta}F^{\alpha\beta}+\frac{m^2}{4}A_{\rho}A^{\rho})
\end{equation}
where index contraction is done using $\eta_{\mu\nu}$.

Therefore we see that the Einstein equations

\begin{equation}
R_{\mu\nu}-\frac{1}{2}\eta_{\mu\nu}(R+\Lambda)=\sum_{a}T_{\mu\nu}^{a}
\end{equation}
are reduced to a set of ordinary differential equations in $r$; or, in
the case of scaling solutions, a set of algebraic equations.

%%%%%%%%%%%%%%%%%%%%%%%%%%%%%%%%%%%%%%%%%%%%%%%%%

\section{Appendix B: Rosetta stone relating different nomenclatures for the Bianchi classification}~\label{Appendix:PateraBianchiDict}
In the table \ref{table1} below, we provide a dictionary relating two different common nomenclatures for the classification of
3d real Lie algebras, as given in \cite{Patera} and \cite{ Ryan, Landau}.
\begin{table}[h]
 \centering
\begin{tabular}{|c|c|}
\hline
Algebra in the notation of \cite{Patera} & Algebra in the notation of \cite{Ryan,Landau}\\
\hline
 $3 A_{1}$ & Bianchi I \\
\hline
 $A_{3,1}$ & Bianchi II\\
\hline
  $A_{2} \oplus A_1$ & Bianchi III\\
\hline
 $A_{3,2}$ & Bianchi IV\\
\hline
 $A_{3,3}$ & Bianchi V\\
\hline
%$A_{3,4}$ & Bianchi VI\\
 $A^a_{3,5}$ ($0 < | a | <1$) %& \\
& Bianchi VI\\
%\hline
%$A_{3,6}$ & Bianchi VII$_0$ \\
\hline
$A^a_{3,7}$ ($a > 0$) %&  \\
& Bianchi VII\\
\hline
  $A_{3,8}$ & Bianchi VIII\\
\hline
$A_{3,9}$ & Bianchi IX\\
\hline
\end{tabular}
\caption{Classification of Real 3 dimensional Lie algebras}\label{table1}
\end{table}
\begin{itemize}
% \item The generators as given in \cite{Ryan, Landau} for Bianchi type VII, are linear combinations of those of $A_{3,6},A^a_{3,7}$ as given in \cite{Patera}.
% \item Similarly the generators for Bianchi type VIII, as in \cite{Ryan, Landau}, are linear combinations of those of $A_{3,8}$ as given in \cite{Patera}. 
 \item Bianchi VII$_0$, which we use many times in this paper, is the special limit of 
Bianchi VII.  More precisely, Bianchi VII has nonzero structure constant $C^1_{23} = - C^1_{32} = -1$, $C^2_{13} = -C^2_{31} = 1$ and $C^2_{23} = -C^2_{32} = a$. By setting $a = 0$, Bianchi VII reduces to Bianchi VII$_0$. 
The $a=0$ limit is called $A_{3,6}$ in \cite{Patera}. 
% is the special limit of the $A^a_{3,7}$ in \cite{Patera} with $a  = 0$.   
 \item $A_{3,4}$ in \cite{Patera} is the special limit of the $A^a_{3,5}$ in \cite{Patera} with $a  = -1$.   
\end{itemize}

%%%%%%%%%%%%%%%%%%%%%%%%%%%%%%%%%%%%%%%%%%%%%%%%%

\section{Appendix C: Null Energy Condition for 4d $A_{4,6}$}{\label{Appendix:NEC}}

We need 
\bea
%\label{nullECforA464D}
f_i \ge 0 \quad (i = 1, 2, 4) \quad \,, \quad 
4 f_2 f_4 \ge (f_3)^2  \,. \quad  
\eea
for $f_i$ given by eq.(\ref{f1forA46typein4d}) - (\ref{f4forA46typein4d}).  
We set $\eta_x = 1$ by performing a coordinate re-parameterization.  
Then, with $\eta_r >0$,  $f_i > 0$ $(i = 1, 2, 4)$ gives  
\bea
 {2  \eta_y (2 b (a-b)+1)- 1- \eta_y^2}  \ge 0 \,, \\
 {2  \eta_y (a-b) (a+2 b)+1 -\eta_y^2} \ge 0 \,, \\   
 {2  \eta_y (a-b) (a+2 b)-1+\eta_y^2}  \ge 0 \,.
\eea
Furthermore, by flipping $r \leftrightarrow -r$, we can always make $a > 0$. 
So the horizon is at $r \to \infty$. In order to satisfy the Nernst's law,  
we require $b > 0$.  

Let's set $a=1$.  
Then the above 3 conditions become 
\bea
\label{firstetayb}
 {2  \eta_y (2 b (1-b)+1)- 1- \eta_y^2}  \ge 0 \,, \\
\label{secondetayb}
 {2  \eta_y (1-b) (1+2 b)+1 -\eta_y^2} \ge 0 \,, \\   
\label{thirdetayb} 
 {2  \eta_y (1-b) (1+2 b)-1+\eta_y^2}  \ge 0 \,.
\eea
In the regime where 
\bea
b > 0 \quad \,, \quad \eta_y > 0\,, 
\eea
the first condition (\ref{firstetayb}) gives 
\bea
\label{A46nullEcondition4dimfirst}
1 + 2 b - 2 b^2 - 2 \sqrt{b - 2 b^3 + b^4} \le \eta_y \le 1 + 2 b - 2 b^2 + 2 \sqrt{b - 2 b^3 + b^4} \,.
\eea
The second the third conditions  (\ref{secondetayb}) and  (\ref{thirdetayb}) give respectively 
\bea
\eta_y \le \frac{1}{2} \left(2 + 2 b - 4 b^2 + \sqrt{4 + (2 + 2 b - 4 b^2)^2}\right) \,, \\
\eta_y \ge \frac{1}{2} \left(-2 - 2 b + 4 b^2 +\sqrt{4 + (2 + 2 b - 4 b^2)^2} \right) \,.
\eea
Finally the condition $4 f_2 f_4 \ge (f_3)^2 $ gives additionally  
\bea
\label{A46nullEcondition4dimlast}
 \frac{1}{4} \left(g_1 
- g_2 \right)
\le \eta_y \le \frac{1}{4} \left(g_1 
+ g_2 \right)
\eea
where
\bea
g_1 &=& \sqrt{8 b \left(10 b^3-4 b^2+b+9\right)+41}  -4 b (b+1)-1  \,, \\
g_2 &=& \sqrt{2} \sqrt{-(2   b+1)^2 \left(\sqrt{8 b \left(10 b^3-4 b^2+b+9\right)+41}-12 (b-1)   b-13\right)}   \,.
\eea

We plot the allowed parameter ranges satisfying (\ref{A46nullEcondition4dimfirst}) - (\ref{A46nullEcondition4dimlast}) in Figure 2. 
\begin{figure}
 \begin{center}
   \includegraphics[width=100mm]{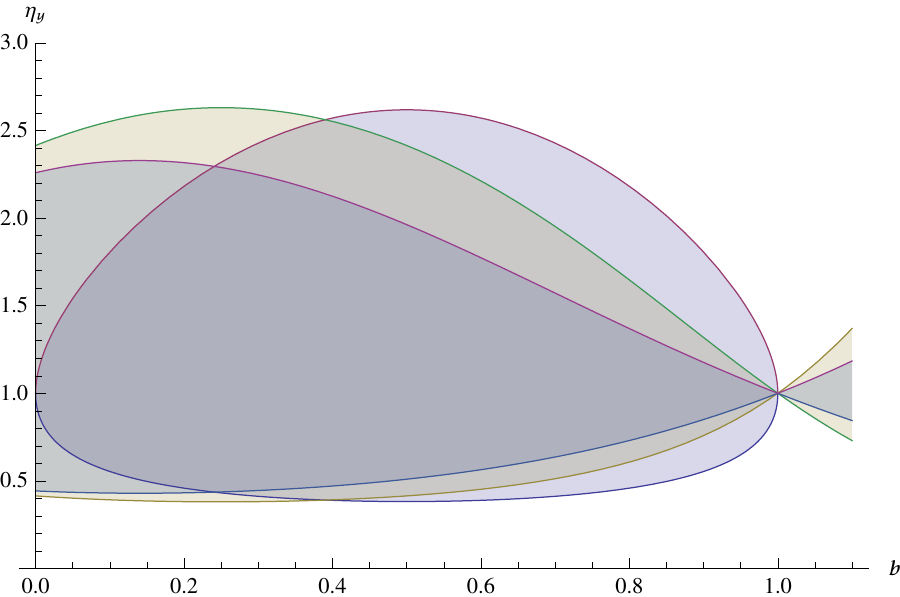}
  \caption{The darkest parameter region is the allowed region in the $(\eta_y , b)$ parameter space. 
  We have set $\eta_x =1$ and $a=1$.}
 \end{center}
\end{figure}
The limit  $\eta_y = b = 1$ corresponds to $AdS_4$. 

%%%%%%%%%%%%%%%%%%%%%%%%%%%%%%%%%%%%%%%%%%%%%%%%%%%%%%%%%%%%%%%%%%%%%%%%%%%%%%%%%%%%%%%%%%%%%%%%%%%

\bibliographystyle{JHEP}
\renewcommand{\refname}{Bibliography}
\addcontentsline{toc}{section}{Bibliography}
\providecommand{\href}[2]{#2}\begingroup\raggedright

\end{document}